\documentclass[a4paper,12pt]{article}
\usepackage[english,russian]{babel}
\usepackage[cp1251]{inputenc}
\usepackage{mathtext}
\usepackage[T2A]{fontenc}
\usepackage{inputenc,amssymb}
\usepackage{epsfig}
\usepackage{caption2}
\usepackage{amsfonts}
\usepackage{amssymb}
\usepackage{amsmath}
\usepackage{geometry}
\usepackage{color}
\usepackage{hyperref}
\hypersetup{
colorlinks=true,
linkcolor=blue,
citecolor=blue,
urlcolor=black
}

\begin{document}

\renewcommand{\r}{\mathbb R}
\newcommand{\eqd}{\stackrel{d}{=}}
\renewcommand{\figurename}{{\bf Figure}}
\renewcommand{\tablename}{{\bf Table}}
\newtheorem{theorem}{Theorem}
\newtheorem{corollary}{Corollary}
\newtheorem{proposition}{Proposition}

\title{\vspace{-1.8cm}
Statistical tests for extreme precipitation volumes}

\author{
V.\,Yu.~Korolev\textsuperscript{1},
A.\,K.~Gorshenin\textsuperscript{2},
K.\,P.~Belyaev\textsuperscript{3}}

\date{}

\maketitle

\footnotetext[1]{Faculty of Computational Mathematics and
Cybernetics, Lomonosov Moscow State University, Russia; Institute of
Informatics Problems, Federal Research Center ``Computer Science and
Control'' of Russian Academy of Sciences, Russia; Hangzhou Dianzi
University, China; \url{vkorolev@cs.msu.su}}

\footnotetext[2]{Institute of Informatics Problems, Federal Research
Center ``Computer Science and Control'' of Russian Academy of
Sciences, Russia; Faculty of Computational Mathematics and
Cybernetics, Lomonosov Moscow State University, Russia; \url{agorshenin@frccsc.ru}}

\footnotetext[3]{P.\,P.~Shirshov Institute of Oceanology of Russian Academy of
Sciences, Russia; Faculty of Computational Mathematics and
Cybernetics, Lomonosov Moscow State University, Russia;
\url{kosbel55@gmail.com}}

{\bf Abstract.} The approaches, based on the negative binomial model for the distribution of
duration of the wet periods measured in days, are proposed to the
definition of extreme precipitation. This model
demonstrates excellent fit with real data and provides a theoretical
base for the determination of asymptotic approximations to the
distributions of the maximum daily precipitation volume within a wet
period as well as the total precipitation volume over a wet period.
The first approach to the definition (and
determination) of extreme precipitation is based on the
tempered Snedecor--Fisher distribution of the maximum daily
precipitation. According to this approach, a daily precipitation
volume is considered to be extreme, if it exceeds a
certain (pre-defined) quantile of the tempered Snedecor--Fisher
distribution. The second approach is based on that the total
precipitation volume for a wet period has the gamma distribution.
Hence, the hypothesis that the total precipitation volume during a
certain wet period is extremely large can be formulated as the
homogeneity hypothesis of a sample from the gamma distribution. Two
equivalent tests are proposed for testing this hypothesis.
Both of these tests deal with
the relative contribution of the total precipitation volume for a
wet period to the considered set (sample) of successive wet periods.
Within the second approach it is possible to introduce the notions
of relatively and absolutely extreme precipitation volumes. 
The results of the application of these tests to real data are
presented yielding the conclusion that the intensity of wet periods
with extreme large precipitation volume increases.

\smallskip

{\bf Key words:} wet periods, total precipitation volume, negative
binomial distribution, asymptotic approximation, extreme order
statistic, random sample size, gamma distribution, Beta
distribution, Snedecor--Fisher distribution, testing statistical
hypotheses.


\section{Introduction}
\label{Introduction}

Estimates of regularities and trends in heavy and extreme daily
precipitation are important for understanding climate variability
and change at relatively small or medium time horizons~\cite{Lockhoff2014}. However,
such estimates are much more uncertain compared to those derived for
mean precipitation or total precipitation during a wet period~\cite{Zolina2014}. This
uncertainty is due to that, first, estimates of heavy precipitation
depend closely on the accuracy of the daily records; they are more
sensitive to missing values~\cite{Zolinaetal2005,Zolinaetal2009}.
Second, uncertainties in the estimates of heavy and extreme
precipitation are caused by the inadequacy of the mathematical
models used for the corresponding calculations. Third, these
uncertainties are boosted by the lack of reasonable means for the
unambiguous (algorithmic) determination of extreme or anomalouslyly
heavy precipitation amplified by some statistical significance
problems owing to the low occurrence of such events. As a
consequence, continental-scale estimates of the variability and
trends in heavy precipitation based on daily precipitation might
generally agree qualitatively but may exhibit significant
quantitative differences. In~\cite{Zolina2013} a detailed review of
this phenomenon is presented where it is noted that for the European
continent, most results hint at a growing intensity of heavy
precipitation over the last five decades.

At the same time, the climate variability and trends at relatively
large time horizons are of no less importance for long-range
business, say, agricultural projects and forecasting of risks of
water floods, dry spells and other natural disasters. In the present
paper we propose a rather reasonable approach to the unambiguous
(algorithmic) determination of extreme or abnormally heavy daily and
total precipitation within a wet period. 

It is traditionally assumed that the duration of a wet period (the number
of subsequent wet days) follows the geometric distribution (for
example, see~\cite{Zolina2013}). But the sequence of dry and wet days is not only independent, it is also devoid of the Markov property~\cite{Gorshenin2017b}. Our approach introduces the negative binomial model for the duration of wet periods measured in days. This model demonstrates excellent fiting the numbers of successive wet days with  the negative binomial distribution with shape
parameter less than one (see~\cite{Gorshenin2017a, Gulev}). It provides a theoretical base for the determination of asymptotic approximations to the distributions of the maximum daily precipitation volume within a wet period and of the total
precipitation volume for a wet period. The asymptotic distribution
of the maximum daily precipitation volume within a wet period turns
out to be a tempered Snedecor--Fisher distribution whereas the total
precipitation volume for a wet period turns out to be the gamma
distribution. Both approximations appear to be very accurate. These
asymptotic approximations are deduced using limit theorems for
statistics constructed from samples with random sizes.

In this paper, two approaches are proposed to the definition of
anomalously extremal precipitation. The first approach to the
definition (and determination) of abnormally heavy daily
precipitation is based on the tempered Snedecor--Fisher
distribution. The second approach is based on the assumption that
the total precipitation volume over a wet period has the gamma
distribution. This assumption is theoretically justified by a
version of the law of large numbers for sums of a random number of
random variables in which the number of summands has the negative
binomial distribution and is empirically substantiated by the
statistical analysis of real data. Hence, the hypothesis that the
total precipitation volume during a certain wet period is anomalously
large can be formulated as the homogeneity hypothesis of a sample
from the gamma distribution. Two equivalent tests are proposed for
testing this hypothesis. One of them is based on the beta
distribution whereas the second is based on the Snedecor--Fisher
distribution. Both of these tests deal with the relative
contribution of the total precipitation volume for a wet period to
the considered set (sample) of successive wet periods. Within the
second approach it is possible to introduce the notions of
relatively abnormal and absolutely anomalous precipitation volumes.
The results of the application of these tests to real data are
presented yielding the conclusion that the intensity of wet periods
with anomalously large precipitation volume increases.

The proposed approaches are to a great extent devoid of the
drawbacks mentioned above: first, estimates of total precipitation
are weakly affected by the accuracy of the daily records and are
less sensitive to missing values. Second, they are based on limit
theorems of probability theorems that yield unambiguous asymptotic
approximations which are used as adequate mathematical models.
Third, these approaches provide unambiguous algorithms for the
determination of extreme or anomalously heavy daily or total
precipitation that do not involve statistical significance problems
owing to the low occurrence of such (relatively rare) events.

Our approaches improve the one proposed in~\cite{Zolinaetal2009},
where an estimate of the fractional contribution from the wettest
days to the total was developed which is less hampered by the
limited number of wet days. For this purpose, in
\cite{Zolinaetal2009} an assumption was enacted (without any theoretical justification) that the statistical regularities in
daily precipitation follow the gamma distribution and the parameters
of the gamma distribution are estimated from the observations. This
assumption made it possible to derive a theoretical distribution of
the fractional contribution of any percentage of wet days to the
total from the gamma distribution function.

The fitted Pareto model for the daily precipitation volume~\cite{GorsheninKorolev2018} together with the observation that the duration of a wet period has the negative
binomial distribution makes it possible to propose a reasonable
model for the distribution of the maximum daily precipitation within
a wet period as an asymptotic approximation provided by the limit
theorems for extreme order statistics in samples with random size.
We will give a strict derivation of such a model having the
form of the tempered Snedecor--Fisher distribution (that is, the
distribution of a positive power of a random variable with the
Snedecor--Fisher distribution) and discuss its properties as well as
some methods of statistical estimation of its parameters. This model
makes it possible to propose the following approach to the
definition (and determination) of an anomalously heavy daily
precipitation volume. The grounds for this approach is an obvious
observation that if $X_1,X_2,\ldots,X_N$ is a sample of $N$ positive
observations, then with finite (possibly, random) $N$, among $X_i$'s
there is {\em always} an extreme observation, say, $X_1$, such that
$X_1\geqslant X_i$, $i=1,2,\ldots,N$. Two cases are possible: (i) $X_1$ is
a `typical' observation and its extreme character is conditioned by
purely stochastic circumstances (there {\em must} be an extreme
observation within a finite homogeneous sample) and (ii) $X_1$ is
abnormally large so that it is an `outlier' and its extreme
character is due to some exogenous factors. It will be shown
that the distribution of $X_1$ in the `typical' case (i) is
the tempered Snedecor--Fisher distribution. Therefore, if $X_1$
exceeds a certain (pre-defined) quantile of the tempered
f distribution (say, of the orders 0.99, 0.995 or
0.999), then it is regarded as `suspicious' to be an outlier, that
is, to be anomalously large (the quantile orders specified above mean
that it is pre-determined that one out of a hundred of maximum daily
precipitations, one out of five hundred of maximum daily
precipitations, or one out of a thousand of maximum daily
precipitations is abnormally large, respectively).

Methodically, this approach is similar to the classical techniques
of dealing with extreme observations~\cite{Embrechts}. The novelty
of the proposed method
is in a more accurate
specification of the distribution of extreme daily precipitation. In
applied problems dealing with extreme values there is a common
tradition which, possibly, has already become a prejudice, that
statistical regularities in the behavior of extreme values
necessarily obey one of well-known three types of extreme value
distributions. In general, this is certainly so, if the sample size
is very large, that is, the time horizon under consideration is very
wide. In other words, the models based on the extreme value
distributions have {\em asymptotic} character. However, in real
practice, when the sample size is finite and the extreme values of
the process under consideration are studied on the time horizon of a
moderate length, the classical extreme value distributions may turn
out to be inadequate models. In these situations a more thorough
analysis may generate other models which appear to be considerably
more adequate. This is exactly the case discussed in the present
paper. Here, within the first approach, along with the `large'
parameter, the expected sample size, one more `small' parameter is
introduced and new models are proposed as asymptotic approximations
when the small parameter is infinitesimal. These models prove to be
exceptionally accurate and demonstrate excellent fit with the
observed data.

To construct another test for distinguishing between the cases (i)
and (ii) mentioned above,
we also strongly improve the results of~\cite{Zolina2013} by giving theoretical grounds for the correct application of the gamma distribution as the model of
statistical regularities of {\em total precipitation volume during a
wet period}. These grounds are based on the negative binomial model
for the distribution of the duration of a wet period. In turn, the
adequacy of the negative binomial model has serious empirical and
theoretical rationale the details of which are described below. With
some caveats the gamma model can be also used for the {\it
conditional} distribution of daily precipitation volumes. The proof
of this result is based on the law of large numbers for random sums
in which the number of summands has the negative binomial
distribution. Hence, the hypothesis that the total precipitation
volume during a certain wet period is anomalously large can be
re-formulated as the homogeneity hypothesis of a sample from the
gamma distribution. Two equivalent statistics are proposed for
testing this hypothesis. The corresponding tests are scale-free and
depend only on the easily estimated shape parameter of the negative
binomial distribution and the time-scale parameter determining the
denominator in the fractional contribution of a wet period under
consideration. It is worth noting that within the second approach
the test for a total precipitation volume during one wet period to
be abnormally large can be applied to the observed time series in a
moving mode. For this purpose a {\em window} (a set of successive
observations) is determined. The observations within a window
constitute the sample to be analyzed. Let $m$ be the number of
observation in the window (the sample size). As the window moves
rightward, each fixed observation falls in exactly $m$ successive
windows (from $m$th to $N-m+1$, where $N$ denotes the number of wet periods). A fixed observation may be recognized as anomalously large
within {\em each} of $m$ windows containing this observation. In
this case this observation will be called {\em absolutely abnormally
large} with respect to a given time horizon (determined by the
sample size $m$. Also, a fixed observation may be recognized as
anomalously large within {\em at least one} of $m$ windows containing
this observation. In this case the observation will be called {\it
relatively abnormally large} with respect to a given time horizon.

The preconditions and backgrounds of all the approaches as well as
their peculiarities will also be discussed.
The main goals of this study are: (i) to introduce the negative binomial distribution as a model distribution to describe the random duration of a wet period and (ii) to show that this model extends the previously used models and better fits to the real observations. Beside that, this paper proves that the (iii) relation of the unique precipitation volume divided by the total precipitation volume taken over the wet period is given by the Snedecor--Fisher distribution and (iv) may be used as a statistical test to estimate the extreme precipitations. This statement also generalizes the previously obtained results from~\cite{Zolinaetal2009}. Finally, the current paper demonstrates that (v) the proposed schemes perfectly fit to the real data.

The paper is organized as follows.
In Section~\ref{SecDailyTest}
we introduce the test for a {\em daily} precipitation volume to be
abnormally large. In Section~\ref{SecTempered} an asymptotic approximation is
proposed for the distribution of the maximum daily precipitation
volume within a wet period. Some analytic properties of the obtained
limit distribution are described. 
Section~\ref{SecFitting} contains the
results and discussion of fitting the distribution proposed in
Section~\ref{SecTempered} to real data. The results of application of the test for
a daily precipitation to be anomalously large based on the tempered
Snedecor--Fisher distribution to real daily precipitation data are
presented and discussed in Section~\ref{SecAnalysis}. Section~\ref{SecTotalTest} deals with the test for a {\em total} precipitation volume over a wet period to be
abnormally large based on testing the homogeneity hypothesis of a
sample from the gamma distribution. Two equivalent statistical tests based on Snedecor--Fisher and beta distributions are
introduced in Section~\ref{SecFisher}. In Section~\ref{SecAbnorm} the application of these tests to a time series in a moving mode is discussed and the notions of
relatively anomalously large and absolutely abnormally large
precipitation are given. The results of application of these
tests to real daily precipitation data are presented and discussed
in Section~\ref{SecAnalysysVol}. Section~\ref{SecConclusion} is devoted to the main conclusions of the work.

\section{The test for a daily precipitation volume to be anomalously large based on the tempered Snedecor--Fisher distribution}
\label{SecDailyTest}

At the beginning of this section we introduce some notation that will be used below. All the r.v.'s under consideration are defined on  the same probability space
$(\Omega,\,\mathfrak{F},\,{\mathbb P})$. The results are expounded in terms of r.v.'s with the corresponding distributions. The symbol $\eqd$ denotes the coincidence of distributions. 

Let $G_{r,\lambda}$ be a r.v. having the gamma distribution with shape parameter $r>0$ and scale parameter $\lambda>0$, that is:
\begin{equation*}
{\mathbb P}(G_{r,\lambda}<x)=\int\limits_{0}^{x}\frac{\lambda^r}{\Gamma(r)}z^{r-1}e^{-\lambda z}dz, \quad x\geqslant 0,
\end{equation*}

Let $W_{\gamma}$ be a r.v. with the Weibull distribution
with the distribution function (d.f.) $\big[1-e^{-x^{\gamma}}\big]{\bf 1}(x\geqslant 0)$ (${\bf 1}(A)$ is the indicator function of a set $A$). The distribution of the r.v. $|X|$, where $X$ is a r.v. with the standard normal d.f.,  is a folded normal ($x\geqslant 0$), that is:
\begin{equation}
\label{FoldedNorm}
{\mathbf P}(|X|<x)=2\Phi(x)-1.
\end{equation}

Let $S_{\alpha,1}$ and $S'_{\alpha,1}$ ($0<\alpha<1$) be i.i.d. r.v.'s with the same strictly stable distribution~\cite{Zolotarev1983}. So, the density $v_{\alpha}(x)$ of
the r.v. $R_{\alpha}=S_{\alpha,1}/S'_{\alpha,1}$ can be represented~\cite{KorolevZeifman2016b,KotzOstrovskii1996} as follows ($x>0$):
\begin{equation}
\label{Rdensity}
v_{\alpha}(x)=\frac{\sin(\pi\alpha)x^{\alpha-1}}{\pi[1+x^{2\alpha}+2x^{\alpha}\cos(\pi\alpha)]}.
\end{equation}

\subsection{The tempered Snedecor--Fisher distribution as an asymptotic
approximation to the maximum daily precipitation volume within a wet period}
\label{SecTempered}

As it has been demonstrated in~\cite{GorsheninKorolev2018,KorolevGorshenin2017b}, the asymptotic probability distribution of extremal daily precipitation within a wet period can be represented as follows (here $r>0$, $\lambda>0$, and $\gamma>0$):
\begin{equation}
\label{Tempered}
F(x; r,\lambda,\gamma)=\bigg(\frac{\lambda x^{\gamma}}{1+\lambda
x^{\gamma}}\bigg)^r,\ \ \ x\geqslant 0.
\end{equation}

Moreover, the theoretical conditions of limit theorems correspond with the real data (in sense of fitting Pareto distribution, see~\cite{GorsheninKorolev2018}). The function~\eqref{Tempered} is a scale mixture of the Fr{\'e}chet (inverse
Weibull) distribution. It can be demonstrated~\cite{GorsheninKorolev2018} for a r.v. $M_{r,\gamma,\lambda}$ with a d.f. $F(x; r,\lambda,\gamma)$ that
\begin{equation*} 
M_{r,\gamma,\lambda}\eqd\Big(\frac{Q_{r,1}}{\lambda
r}\Big)^{1/\gamma}.
\end{equation*}

That is, the distribution of the r.v. $M_{r,\gamma,\lambda}$ up to a
non-random scale factor coincides with that of the positive power of
a r.v. with the Snedecor--Fisher distribution. In other words, the
distribution function $F(x; r,\lambda,\gamma)$~\eqref{Tempered} up to a power
transformation of the argument $x$ coincides with the
Snedecor--Fisher distribution function. In statistics, distributions
with arguments subjected to the power transformation are
conventionally called {\em tempered}. Therefore, we have serious
reason to call the distribution $F(x; r,\lambda,\gamma)$ {\it
tempered Snedecor--Fisher distribution}. Some properties of the distribution of the r.v.
$M_{r,\gamma,\lambda}$ were discussed in
\cite{GorsheninKorolev2018}. 
In particular, it was shown that the limit distribution~\eqref{Tempered} can be
represented as a scale mixture of exponential or stable or Weibull
or Pareto or folded normal laws ($r\in(0,1]$, $\gamma\in(0,1]$, $\lambda>0$):
\begin{equation*} 
M_{r,\gamma,\lambda}\eqd
\frac{G_{r,\lambda}^{1/\gamma}S_{\gamma,1}}{W_1}\eqd
\frac{W_{\gamma}}{W'_{\gamma}}\cdot\frac{1}{Z_{r,\lambda}^{1/\gamma}}\eqd
W_1\cdot\frac{R_{\gamma}}{W'_1Z_{r,\lambda}^{1/\gamma}}\eqd
\frac{\Pi R_{\gamma}}{Z_{r,\lambda}^{1/\gamma}}\eqd
\frac{|X|\sqrt{2W_1}R_{\gamma}}{W'_1Z_{r,\lambda}^{1/\gamma}},
\end{equation*}
where $W_{\gamma}\eqd W'_{\gamma}$, $W_1\eqd W'_1$, the r.v.
$R_{\gamma}$ has the density {\rm\eqref{Rdensity}}, the r.v. $\Pi$
has the Pareto distribution (${\mathbb P}(\Pi>x)=(x+1)^{-1}$, $x\geqslant 0$),
and in each term the involved r.v$:$s are independent. 

It should be mentioned that the same mathematical reasoning can be
used for the determination of the asymptotic distribution of the
maximum daily precipitation within $m$ wet periods with arbitrary
finite $m\in\mathbb{N}$. Indeed, fix arbitrary positive
$r_1,\ldots,r_m$ and $p\in(0,1)$. Let
$N_{r_1,p}^{(1)},\ldots,N_{r_m,p}^{(m)}$ be independent random
variables having the negative binomial distributions with parameters
$r_j,\,p$, $j=1,\ldots,m$, respectively. By the consideration of
characteristic functions it can be easily verified that
\begin{equation}\label{NBsum}
N_{r_1,p}^{(1)}+\ldots+N_{r_m,p}^{(m)}\eqd N_{r,p},
\end{equation}
where $r=r_1+\ldots+r_m$. If all $r_j$ coincide, then $r=mr_1$ and
in accordance with the results of papers~\cite{GorsheninKorolev2018,KorolevGorshenin2017b} and relation~\eqref{NBsum}, the asymptotic
distribution of the maximum daily precipitation within $m$ wet
periods has the form ($x\geqslant 0$)
\begin{equation*}
F^{(m)}(x; r,\lambda,\gamma)=F(x; mr_1,\lambda,\gamma)=
\bigg(\frac{\lambda x^{\gamma}}{1+\lambda x^{\gamma}}\bigg)^{mr_1}.
\end{equation*}

And if now $m$ infinitely increases and simultaneously $\lambda$
changes as $\lambda=cm$, $c\in(0,\infty)$, then, obviously,
\begin{equation*}
\lim_{m\to\infty}F^{(m)}(x; r,\lambda,\gamma)=\lim_{m\to\infty}F(x;
mr_1,cm,\gamma)=\lim_{n\to\infty}\Big(1-\frac{1}{1+cmx^{\gamma}}\Big)^{mr_1}=e^{-\mu
x^{-\gamma}}
\end{equation*}
with $\mu=(cr_1)^{-1}$, that is, the distribution function
$F^{(m)}(x; r,\lambda,\gamma)$ of the maximum daily precipitation
within $m$ wet periods turns into the classical Fr{\'e}chet distribution. 

\subsection{The algorithms of statistical fitting of the tempered Snedecor--Fisher distribution model to real data}
\label{SecFitting}

Some methods of statistical estimation of the parameters $r$, $\lambda$ and $\gamma$ of the tempered Snedecor--Fisher distribution~\eqref{Tempered}
were described in~\cite{GorsheninKorolev2018}. In this section the algorithms and corresponding formulas for practical computation are briefly given.

Let $\{X_{i,j}\}$, $i=1,\ldots,m$, $j=1,\ldots,m_i$, be the precipitation
volumes on the $j$th day of the $i$th wet sequence.
\begin{equation}
X^*_k=\max\{X_{k,1},\ldots,X_{k,m_k}\},\ \ \
k=1,\ldots,m.\label{VarSample}
\end{equation}
Let $X^*_{(1)},\ldots,X^*_{(m)}$ be order statistics constructed
from the sample $X^*_1,\ldots,X^*_m$, where $X^*_k=\max\{X_{k,1},\ldots,X_{k,m_k}\}$. The unknown parameters $r$, $\lambda$ and $\gamma$ can be found as a solution of a following system of equations  (for fixed values $p_1$, $p_2$ and $p_3$, $0<p_1<p_2<p_3<1$):
\begin{equation*}
X^*_{([mp_k])}=\left(\frac{p_k^{1/r}}{\lambda-\lambda p_k^{1/r}}\right)^{1/\gamma}, \quad k=1,2,3
\end{equation*}
(here the symbol $[a]$ denotes the integer part of a number $a$).

\begin{proposition}
The values of parameters $\gamma$ and $\lambda$ can be estimated as follows:
\begin{equation}
\widetilde\gamma=\frac{\frac1r(\log p_1-\log
p_3)+\log(1-p_3^{\frac1r})-\log(1-p_1^{\frac1r})}{\log X^*_{([mp_1])}-\log
X^*_{([mp_3])}}, \label{gamma}
\end{equation}
\begin{equation}
\widetilde\lambda=\frac{p_2^{\frac1r}}{(1-p_2^{\frac1r})(X^*_{([mp_2])})^{\gamma}}.
\label{lambda}
\end{equation}
\end{proposition}



\begin{proposition}
If the value of parameter $r$ is estimated as a corresponding parameter of the negative
binomial distribution, least squares estimates of parameters $\gamma$ and $\lambda$ are as follows:
\begin{equation}\label{gammaLS}
\widehat\gamma=\frac{(m-1)\sum\limits_{i=1}^{m-1}\log\frac{i^{1/r}}{m^{1/r}-i^{1/r}}\log
X^*_{(i)}-\sum\limits_{i=1}^{m-1}\log
X^*_{(i)}\sum\limits_{i=1}^{m-1}\log\frac{i^{1/r}}{m^{1/r}-i^{1/r}}}{(m-1)\sum\limits_{i=1}^{m-1}(\log
X^*_{(i)})^2-(\sum\limits_{i=1}^{m-1}\log X^*_{(i)})^2},
\end{equation}
\begin{equation}\label{lambdaLS}
\widehat\lambda=\exp\Big\{\frac{1}{m-1}\Big(\sum\nolimits_{i=1}^{m-1}\log\frac{i^{1/r}}{m^{1/r}-i^{1/r}}-\widehat\gamma\sum\nolimits_{i=1}^{m-1}\log
X^*_{(i)}\Big)\Big\}.
\end{equation}
\end{proposition}

The numerical results of estimation of the parameters of daily precipitation in Potsdam and Elista from $1950$ to $2009$ using both algorithms are presented in Tables~\ref{TabPotsdam} and~\ref{TabElista}. The
first column indicates the censoring threshold: since the tempered
Snedecor--Fisher distribution is an asymptotic model which is
assumed to be more adequate with small ``success probability'', the
estimates were constructed from differently censored samples which
contain only those wet periods whose duration is no less than the
specified threshold. The second column contains the correspondingly
censored sample size. The third and fourth columns contain the
sup-norm discrepancy between the empirical and fitted tempered
Snedecor--Fisher distribution for two types of estimators (quantile
and least squares) described above. The rest columns contain the
corresponding values of the parameters estimated by these two
methods. According to Tables~\ref{TabPotsdam} and~\ref{TabElista},
the best accuracy is attained when the censoring threshold equals $3$
days for Elista and $5$--$6$ days for Potsdam. The least squares method~\eqref{gammaLS} and~\eqref{lambdaLS} leads to the more accurate estimates. The vivid examples of approximation of the real data with the functions $F(x;r,\gamma,\lambda)$ are presented in~\cite{GorsheninKorolev2018}). The corresponding numerical methods have been implemented using MATLAB built-in programming language.

\begin{table}[!h]
\caption{Potsdam ($r=0.847$)}\label{TabPotsdam} \centering
\begin{tabular}{cccccccc}
\hline\noalign{\smallskip}
{Minimum}&{Sample}&{Discrepancy,}&{Discrepancy,}&$\widetilde\lambda$&$\widehat\lambda$&$\widetilde\gamma$&$\widehat\gamma$\\
{duration}&{size}&{quantile metod:}&{LS
method:} &{quantile}&
{LS}&{quantile}&{LS}\\
& &{\eqref{gamma},~\eqref{lambda}}&{\eqref{gammaLS},
\eqref{lambdaLS}}&{method}&{method}&{method}&{method}\\
\noalign{\smallskip}\hline\noalign{\smallskip}
$1$&$3323$ &$0.09$&$0.092$&$0.169$&$0.211$&$1.177$&$1.29$\\
$2$&$2066$ &$0.045$&$0.065$&$0.0381$&$0.0538$&$1.76$&$1.709$\\
$3$&$1282$ &$0.031$&$0.041$&$0.01$&$0.013$&$2.261$&$2.189$\\
$4$&$862$ &$0.026$&$0.027$&$0.00487$&$0.00454$&$2.449$&$2.523$\\
$6$&$384$ &$0.025$&$0.026$&$0.0016$&$0.0012$&$2.822$&$2.948$\\
$8$&$163$ &$0.04$&$0.045$&$0.0007$&$0.0005$&$3.174$&$3.253$\\
$10$&$73$ &$0.041$&$0.042$&$0.0003$&$0.0003$&$3.389$&$3.352$\\
$15$&$12$ &$0.13$&$0.09$&$0.0014$&$0.0009$&$2.667$&$2.972$\\
\noalign{\smallskip}\hline
\end{tabular}
\end{table}

\begin{table}[!h]
\caption{Elista ($r=0.876$)}\label{TabElista} \centering
\begin{tabular}{cccccccc}
\hline\noalign{\smallskip}
{Minimum}&{Sample}&{Discrepancy,}&{Discrepancy,}&$\widetilde\lambda$&$\widehat\lambda$&$\widetilde\gamma$&$\widehat\gamma$\\
{duration}&{size}&{quantile metod:}&{LS
method:} &{quantile}&
{LS}&{quantile}&{LS}\\
& &{\eqref{gamma},~\eqref{lambda}}&{\eqref{gammaLS},
\eqref{lambdaLS}}&{method}&{method}&{method}&{method}\\
\noalign{\smallskip}\hline\noalign{\smallskip}
$1$&$2937$&$0.06$&$0.06$&$0.361$&$0.347$&$1.057$&$1.266$\\
$2$&$1374$&$0.049$&$0.055$&$0.108$&$0.1$&$1.42$&$1.576$\\
$3$&$656$&$0.041$&$0.045$&$0.0454$&$0.0377$&$1.706$&$1.898$\\
$4$&$319$&$0.051$&$0.06$&$0.0231$&$0.0272$&$1.899$&$1.94$\\
$6$&$77$&$0.07$&$0.075$&$0.0178$&$0.0144$&$2.017$&$2.186$\\
$7$&$42$&$0.15$&$0.01$&$0.0201$&$0.0206$&$1.974$&$2.184$\\
$8$&$22$&$0.12$&$0.14$&$0.0143$&$0.0355$&$2.003$&$1.769$\\
$10$&$10$&$0.17$&$0.16$&$0.0137$&$0.0377$&$2.154$&$1.798$\\
\noalign{\smallskip}\hline
\end{tabular}
\end{table}

\subsection{The examples of statistical analysis of daily precipitation}
\label{SecAnalysis}

The approach to the determination of an anomalously heavy daily
precipitation is methodically similar to the classical techniques of
dealing with extreme observations~\cite{Embrechts}. The novelty of
the proposed method is in an accurate specification of the
mathematical model of the distribution of extreme daily
precipitation which turned out to be the tempered Snedecor--Fisher
distribution.

The algorithm of determination of an anomalously heavy daily
precipitation is as follows. First, the parameters of the
distribution function $F(x; r,\lambda,\gamma)$ are estimated from
the historical data. Second, a small positive number $\varepsilon$
is fixed. Third, the $(1-\varepsilon)$-quantile
$\tau(1-\varepsilon;\,r,\lambda,\gamma)$ of the distribution
function $F(x; r,\lambda,\gamma)$ is calculated. If the maximum value, say, $X$ of the daily precipitation volume
within some wet period exceeds
$\tau(1-\varepsilon;\,r,\lambda,\gamma)$, then $X$ is regarded as
`suspicious' to be an outlier, that is, to be anomalously large.
It is easy to see that the the probability of the error of the first
kind (occurring in the case where a `regularly large' maximum value
is erroneously recognized as an anomalously large outlier) for this
test is approximately equal to $\varepsilon$.

\begin{figure}[!h]
\begin{minipage}[h]{\textwidth}
\center{\includegraphics[width=\textwidth, height=0.3\textheight]{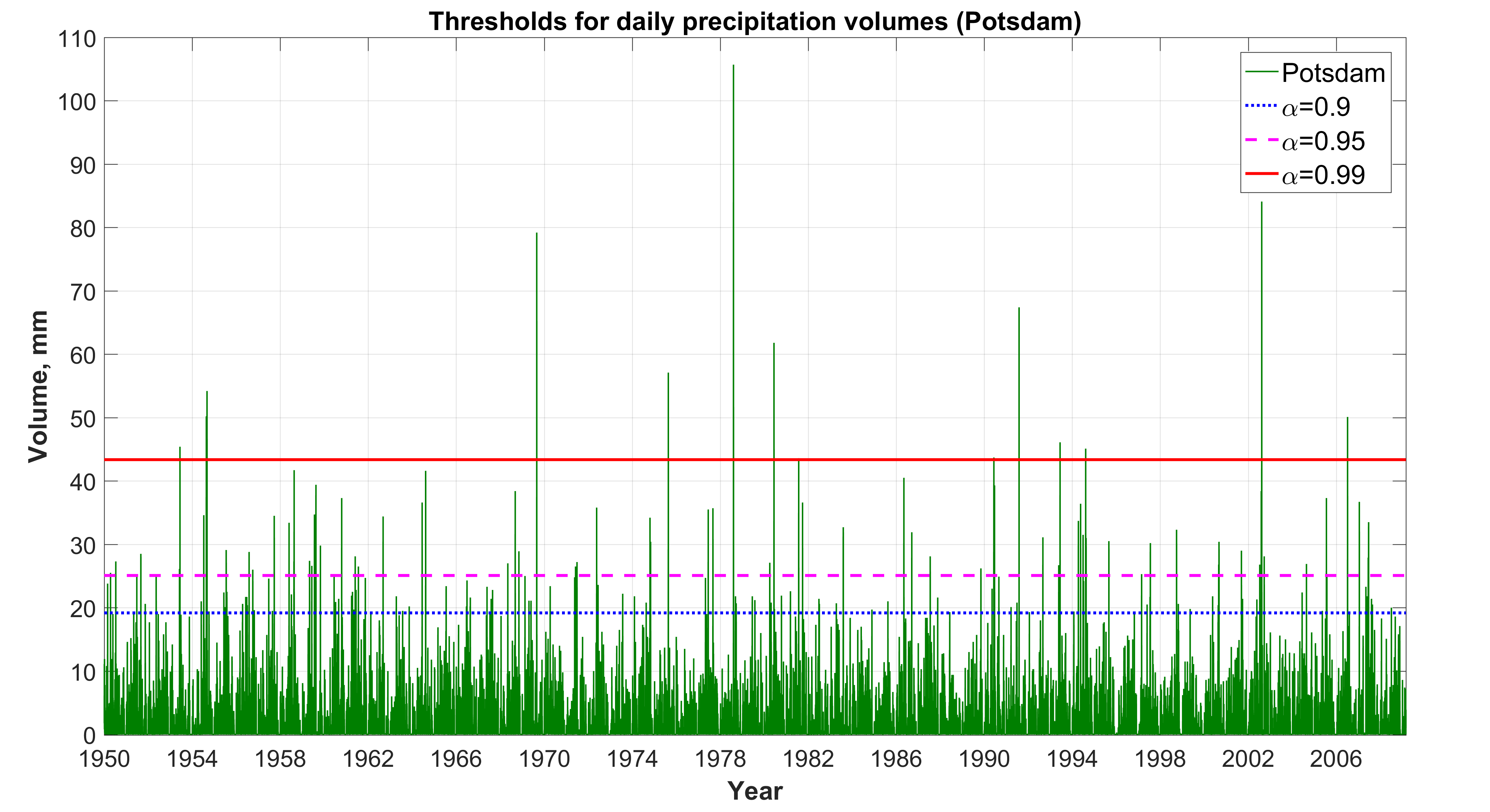}
\\a)}
\end{minipage}
\vfill
\begin{minipage}[h]{\textwidth}
\center{\includegraphics[width=\textwidth,
height=0.3\textheight]{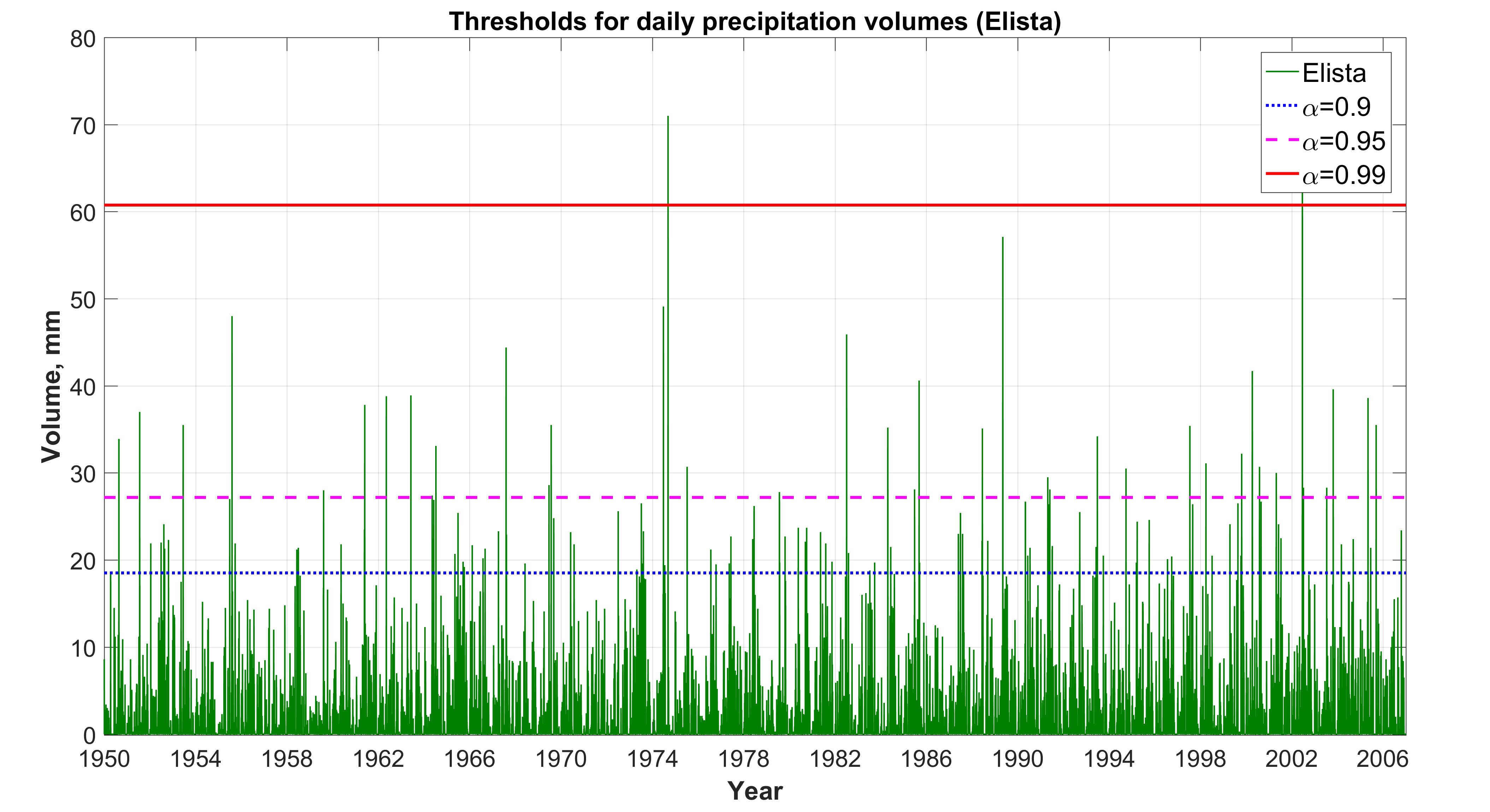}
\\ b)}
\end{minipage}
\caption{Testing maximum daily precipitation within a wet period for
abnormal heaviness: a) Potsdam; b) Elista, all data.}
\label{FigDailyMax1}
\end{figure}

\begin{figure}[!h]
\begin{minipage}[!h]{\textwidth}
\center{\includegraphics[width=\textwidth,
height=0.3\textheight]{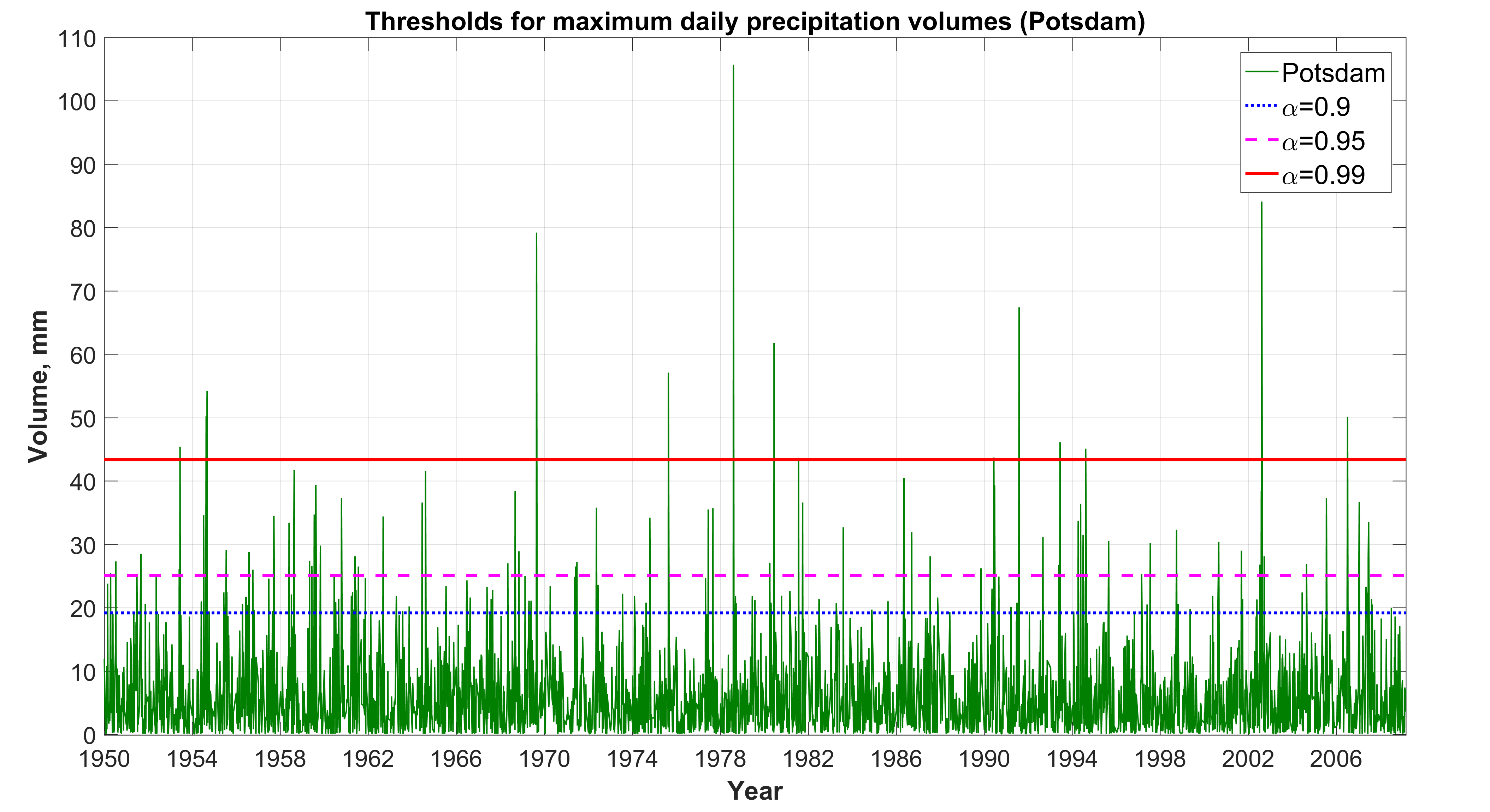}
\\a)}
\end{minipage}
\vfill
\begin{minipage}[!h]{\textwidth}
\center{\includegraphics[width=\textwidth,
height=0.3\textheight]{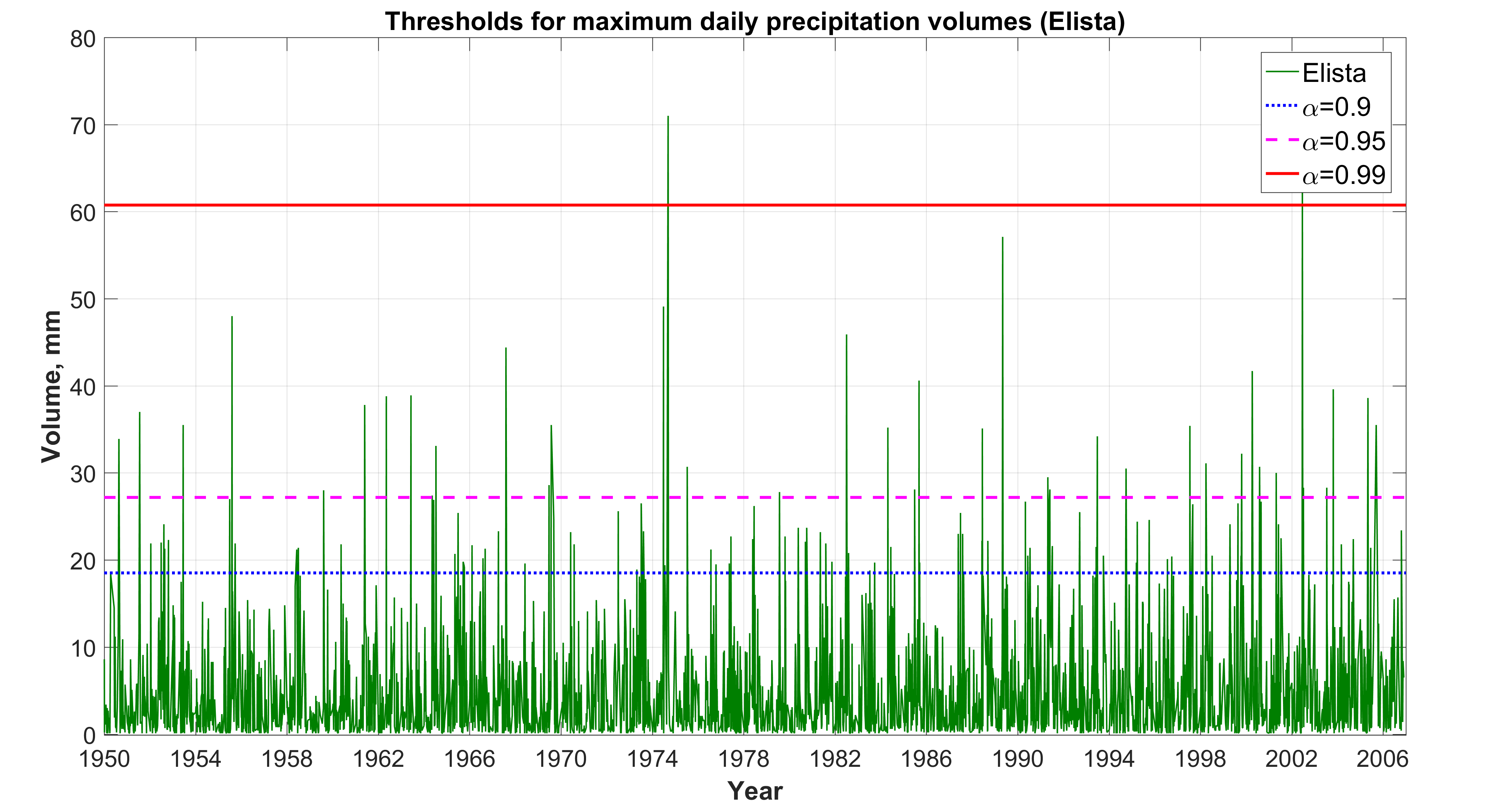}
\\ b)}
\end{minipage}
\caption{Testing maximum daily precipitation within a wet period for
abnormal heaviness: a) Potsdam; b) Elista, data containing only
maximum daily precipitation for every wet period.}
\label{FigDailyMax2}
\end{figure}

The application of this test to real data is illustrated by Figs.~\ref{FigDailyMax1} and~\ref{FigDailyMax2}. On these figures the
lower horizontal line corresponds to the threshold equal to the
quantile of the fitted tempered Snedecor--Fisher distribution of
order $0.9$. The middle and upper lines correspond to the quantiles of
orders $0.95$ and $0.99$, respectively.
 

Fig.~\ref{FigDailyMax1} contains all data. For the sake of
vividness, on Fig.~\ref{FigDailyMax2} only one, maximum, daily
precipitation is exposed for each wet period. From Fig.~\ref{FigDailyMax2} it is seen that during $58$ years (from $1950$ to
$2007$) in Potsdam there were $13$ wet periods containing anomalously
heavy maximum daily precipitation volumes (at $99\%$ threshold) and $69$
wet periods containing anomalously heavy maximum daily precipitation
volumes (at $95\%$ threshold). Other maxima were `regular'. During the
same period in Elista there were only $2$ wet periods containing
anomalously heavy maximum daily precipitation volumes (at $99\%$
threshold) and $4$0 wet periods containing anomalously heavy maximum
daily precipitation volumes (at $95\%$ threshold). Other maxima were
`regular'. The proportion of abnormal maxima exceeding $99\%$ and $95\%$
thresholds in Potsdam is quite adequate (the latter is approximately
five times greater than the former) whereas in Elista this
proportion is noticeably different. Perhaps, this can be explained
by the fact that, for Elista, heavy rains are rare events.

\section{The tests for a total precipitation volume to be anomalously extremal based on the homogeneity test of a sample from the gamma distribution}
\label{SecTotalTest}

\subsection{The tests based on the beta and Snedecor--Fisher distributions}
\label{SecFisher}

Here we will propose some algorithms of testing the hypotheses that
a {\em total precipitation volume during a wet period} is anomalously
extremal within a certain time horizon. Moreover, our approach makes
it possible to consider relatively anomalously extremal volumes and
absolutely anomalously extremal volumes for a given time horizon.

Let $m\in\mathbb{N}$ and
$G^{(1)}_{r,\mu},G^{(2)}_{r,\mu},\ldots,G^{(m)}_{r,\mu}$ -- be
independent r.v.'s having the same gamma distribution with shape
parameter $r>0$ and scale parameter $\mu>0$. In
\cite{Zolinaetal2009} it was suggested to use the distribution of
the ratio
\begin{equation*} 
R^*=\frac{G^{(1)}_{r,\mu}}{G^{(1)}_{r,\mu}+G^{(2)}_{r,\mu}+\ldots+G^{(m)}_{r,\mu}}
\eqd
\frac{G^{(1)}_{r,1}}{G^{(1)}_{r,1}+G^{(2)}_{r,1}+\ldots+G^{(m)}_{r,1}}
\end{equation*}
as a heuristic model of the distribution of the extremely large
precipitation volume based on the assumption that fluctuations of
{\em daily} precipitation follow the gamma distribution. The gamma model for the distribution of daily precipitation volume is less adequate than the Pareto one~\cite{GorsheninKorolev2018}. Here we
will modify the technique proposed in~\cite{Zolinaetal2009} and make
it more adequate and justified.

\begin{figure}[!h]
\center{\includegraphics[width=0.9\textwidth,
height=0.4\textwidth]{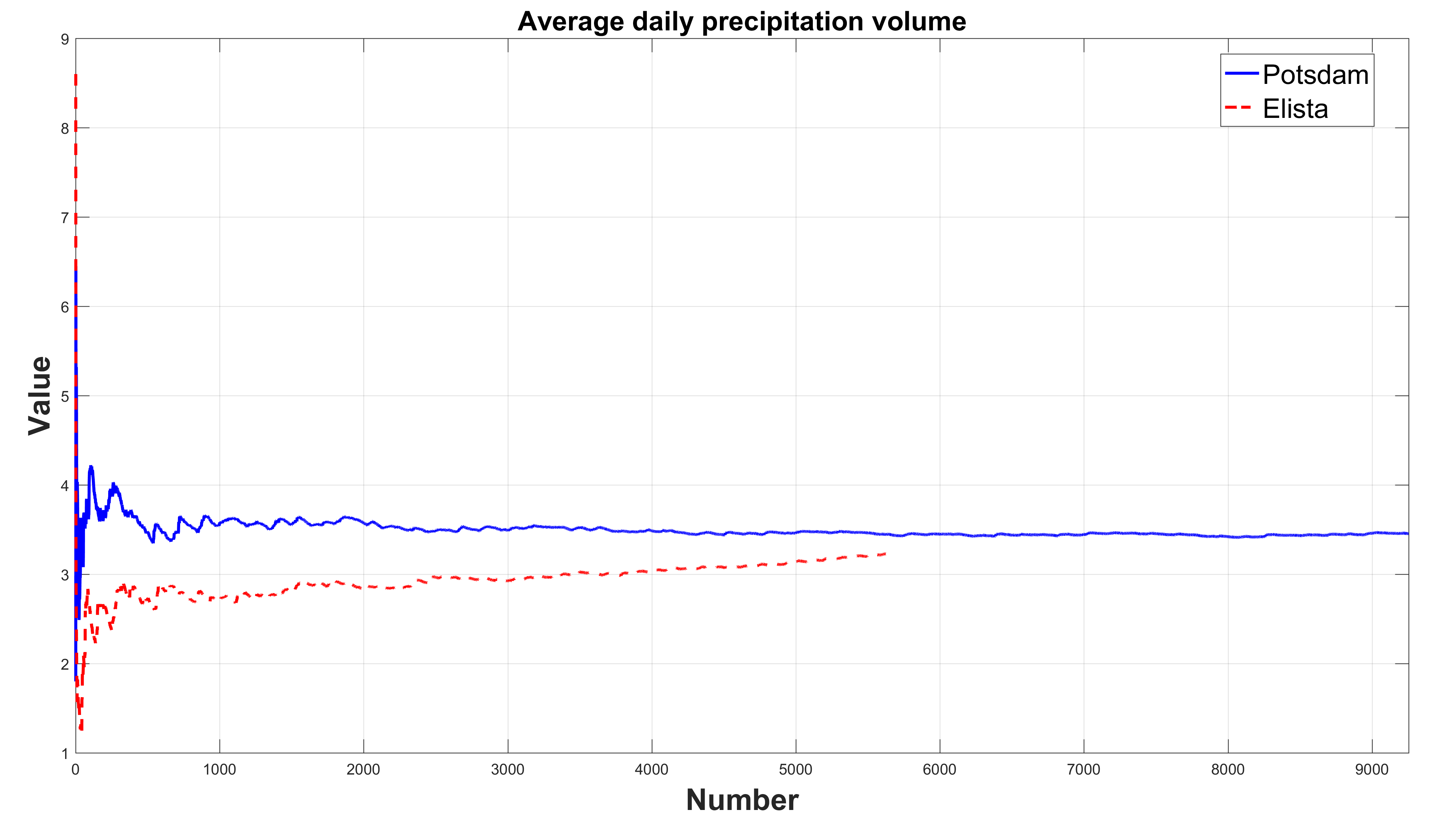}} \caption{Stabilization of the
cumulative averages of daily precipitation volumes as $n$ grows in
Potsdam (continuous line) and Elista (dash line).
\label{FigAvg}}
\end{figure}

Let $X_1,X_2,\ldots$ be daily precipitation volumes on wet days. For
$k\in\mathbb{N}$ denote $S_k=X_1+\ldots+X_k$. The statistical
analysis of the observed data shows that the average daily
precipitation volume on wet days is finite:
\begin{equation}\label{LLN}
\frac{1}{n}\sum\limits_{j=1}^nX_j\Longrightarrow a\in(0,\infty).
\end{equation}
Here the symbol $\Longrightarrow$ denotes the convergence in distribution. 

Fig.~\ref{FigAvg} illustrates the stabilization of the cumulative averages of
daily precipitation volumes as $n$ grows in Potsdam (continuous
line) and Elista (dash line), and thus, the practical validity of
assumption~\eqref{LLN}. It should be emphasized that in~\eqref{LLN} we {\em do not} assume
that $X_1,X_2,\ldots$ are independent.

Let $r>0$, $\mu>0$, $q\in(0,1)$, $n\in\mathbb{N}$. Let the r.v.
$N_{r,p_n}$ have the negative binomial distribution with parameters
$r$ and $p_n=\min\{q,\mu/n\}$. Using the properties of
characteristic functions it is easy to make sure that
\begin{equation}
n^{-1}N_{r,p_n}\Longrightarrow
G_{r,\mu}\eqd{\textstyle\frac1\mu}G_{r,1}\label{Cond}
\end{equation}
as $n\to\infty$. From~\eqref{Cond} and From~\eqref{Cond} and the transfer theorem for random sequences with independent random indices (see \cite{Korolev1994, Korolev1995}) we obtain the following analog of the {\em law of large numbers for negative binomial random sums} which can be actually regarded as a generalization of the
R{\'e}nyi theorem concerning the rarefaction of renewal processes.

\begin{theorem}
Assume that the daily precipitation volumes on
wet days $X_1,X_2,...$ satisfy condition {\rm\eqref{LLN}}. Let the
numbers $r>0$, $q\in(0,1)$ and $\mu>0$ be arbitrary. For each
$n\in\mathbb{N}$, let the r.v. $N_{r,p_n}$ have the negative
binomial distribution with parameters $r$ and $p_n=\min\{q,\mu/n\}$.
Assume that the r.v.'s $N_{r,p_n}$ are independent of the sequence
$X_1,X_2,...$ Then
\begin{equation*}
n^{-1}S_{N_{r,p_n}}\Longrightarrow aG_{r,\mu}\eqd
{\textstyle\frac{a}{\mu}} G_{r,1}
\end{equation*}
as $n\to\infty$. 
\end{theorem}

\begin{figure}[!h]
\begin{minipage}[h]{\textwidth}
\center{\includegraphics[width=\textwidth,
height=0.3\textheight]{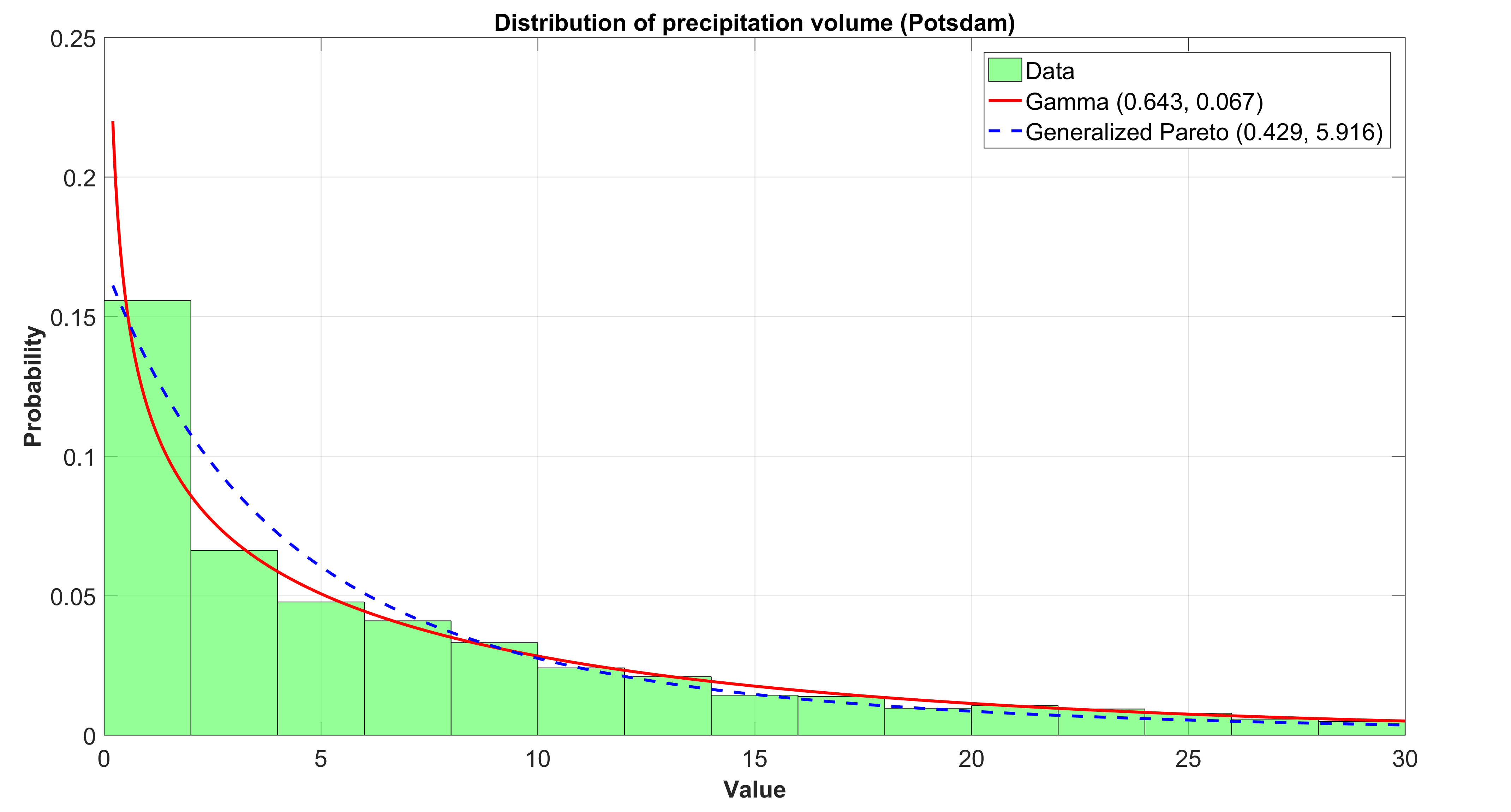}
\\a)}
\end{minipage}
\vfill
\begin{minipage}[h]{\textwidth} \center{
\includegraphics[width=\textwidth,
height=0.3\textheight]{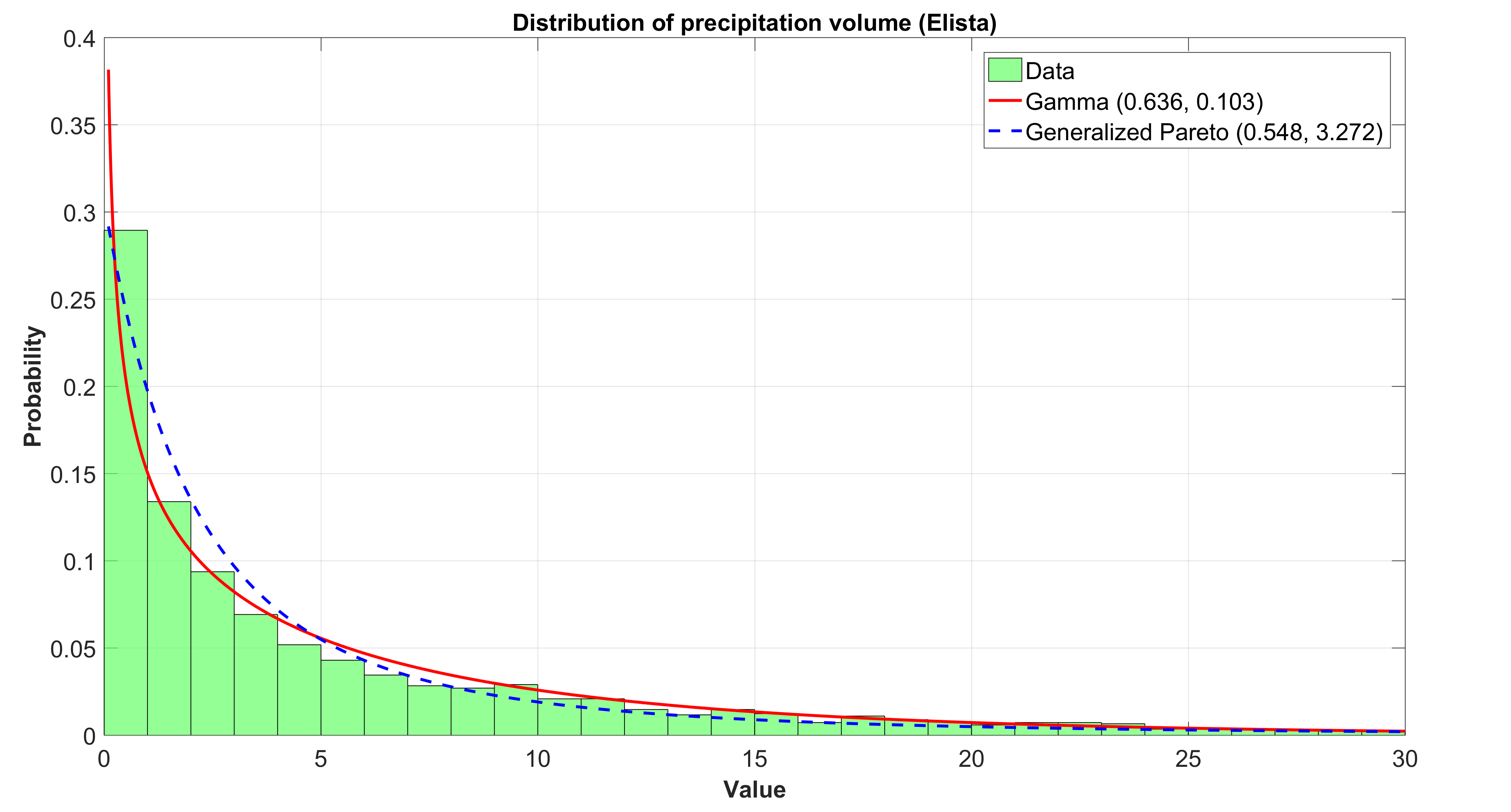} \\ b)}
\end{minipage}
 \caption{The histograms of total
precipitation volumes in Potsdam (a) and Elista (b) and the fitted
gamma and generalized Pareto distributions.} \label{FigGammaTotalPrecip}
\end{figure}

Therefore, with the account of the excellent fit of the negative
binomial model for the duration of a wet period~\cite{GorsheninKorolev2018}, with
rather small $p_n$, the gamma distribution can be regarded as an
adequate and theoretically well-based model for the total
precipitation volume during a (long enough) wet period. This
theoretical conclusion based on the negative binomial model for the
distribution of duration of a wet period is vividly illustrated by
the empirical data as shown on Fig.~\ref{FigGammaTotalPrecip} where
the histograms of total precipitation volumes in Potsdam and
Elista and the fitted gamma distributions are shown. For
comparison, the densities of the best generalized Pareto
distributions are also presented. It can be seen that even the best
fitted Pareto distributions demonstrate worse fit than the gamma
distribution.


Let $m\in\mathbb{N}$ and
$G^{(1)}_{r,\mu},G^{(2)}_{r,\mu},\ldots,G^{(m)}_{r,\mu}$ be
independent r.v.'s having the same gamma distribution with
parameters $r>0$ and $\mu>0$. Consider the relative contribution of the r.v. $G^{(1)}_{r,\mu}$ to
the sum $G^{(1)}_{r,\mu}+G^{(2)}_{r,\mu}+\ldots+G^{(m)}_{r,\mu}$:
\begin{equation*}
R=\frac{G^{(1)}_{r,\mu}}{G^{(1)}_{r,\mu}+G^{(2)}_{r,\mu}+
\ldots+G^{(m)}_{r,\mu}}\eqd\frac{G^{(1)}_{r,1}}{G^{(1)}_{r,1}+G^{(2)}_{r,1}+
\ldots+G^{(m)}_{r,1}}\eqd
\end{equation*}
\begin{equation}\label{R}
\eqd\bigg(1+\frac{1}{G^{(1)}_{r,1}}(G^{(2)}_{r,1}+\ldots+G^{(m)}_{r,1})\bigg)^{-1}\eqd
\bigg(1+\frac{G_{(m-1)r,1}}{G_{r,1}}\bigg)^{-1},
\end{equation}
where the gamma-distributed r.v.'s on the right hand side are
independent. So, the r.v. $R$ characterizes the relative
precipitation volume for one (long enough) wet period with respect
to the total precipitation volume registered for $m$ wet periods.

The distribution of the r.v. $R$ is completely determined by the
distribution of the ratio of two {\em independent} gamma-distributed
r.v.'s. To find the latter, denote $k=(m-1)r$ and obtain
\begin{equation*}
\frac{G_{k,1}}{G_{r,1}}=\frac{k}{r}\cdot\bigg(\frac
rk\cdot\frac{G_{k,1}}{G_{r,1}}\bigg)\eqd\frac{k}{r}\cdot Q_{k,r},
\end{equation*}
where $Q_{k,r}$ is the r.v. having the Snedecor--Fisher distribution
determined for $k>0$, $r>0$ by the Lebesgue density
\begin{equation}
f_{k,r}(x)=\frac{\Gamma(k+r)}{\Gamma(k)\Gamma(r)}
\Big(\frac{k}{r}\Big)^{k}\frac{x^{k-1}}{(1+\frac{k}{r}x)^{k+r}},\ \
\ x\geqslant 0,\label{FisherPDF}
\end{equation}
(as is known, $Q_{k,r}\eqd r G_{k,\,1}(k G_{r,\,1})^{-1}$, where the
r.v.'s $G_{k,\,1}$ and $G_{r,\,1}$ are independent (see, e. g.,
\cite{Johnson}, p. 32)). It should be noted that the particular
value of the scale parameter is insignificant. For convenience, it
is assumed equal to one.

So, $R\eqd\big(1+\frac{k}{r}Q_{k,r}\big)^{-1}$, and, as is easily
made sure by standard calculation using~\eqref{FisherPDF}, the distribution
of the r.v. $R$ is determined by the density
\begin{equation*}
p(x;k,r)=\frac{\Gamma(k+r)}{\Gamma(r)\Gamma(k)}(1-x)^{k-1}x^{r-1},\
\ \ \ 0\leqslant x\leqslant1,
\end{equation*}
that is, it is the beta distribution with parameters $k=(m-1)r$ and
$r$.

Then the test for the homogeneity of an independent sample of size
$m$ consisting of the gamma-distributed observations of total
precipitation volumes during $m$ wet periods with known $\gamma$
based on the r.v. $R$ looks as follows. Let $V_1,\ldots,V_m$ be the
total precipitation volumes during $m$ wet periods and, moreover,
$V_1\geqslant V_j$ for all $j\geqslant 2$. Calculate the quantity
\begin{equation*}
SR=\frac{V_1}{V_1+\ldots+V_m}
\end{equation*}
($SR$ means ``$S$ample $R$''). From what was said above it follows
that under the hypothesis $H_0$: ``the precipitation volume $V_1$
under consideration {\em is not} anomalously large'' the r.v. $SR$
has the beta distribution with parameters $k=(m-1)r$ and $r$. Let
$\varepsilon\in(0,1)$ be a small number,
$\beta_{k,r}(1-\varepsilon)$ be the $(1-\varepsilon)$-quantile of
the beta distribution with parameters $k=(m-1)r$ and $r$. If
$SR>\beta_{k,r}(1-\varepsilon)$, then the hypothesis $H_0$ must be
rejected, that is, the volume $V_1$ of precipitation during one wet
period must be regarded as anomalously large. Moreover, the
probability of erroneous rejection of $H_0$ is equal to
$\varepsilon$.

Instead of $R$~\eqref{R}, the quantity
\begin{equation*}
R_0=\frac{(m-1)G^{(1)}_{r,\mu}}{G^{(2)}_{r,\mu}+
\ldots+G^{(m)}_{r,\mu}}\eqd
\frac{k}{r}\frac{G_{r,\mu}}{G_{k,\mu}}\eqd\frac{k}{r}\frac{G_{r,1}}{G_{k,1}}\eqd
Q_{r,k}
\end{equation*}
can be considered. Then, as is easily seen, the r.v.'s $R$ and $R_0$
are related by the one-to-one correspondence
\begin{equation*}
R=\frac{R_0}{m-1+R_0} \ \ \text{\rm or} \ \ R_0=\frac{(m-1)R}{1-R},
\end{equation*}
so that the homogeneity test for a sample from the gamma
distribution equivalent to the one described above and,
correspondingly, the test for a precipitation volume during a wet
period to be anomalously large, can be based on the r.v. $R_0$ which
has the Snedecor--Fisher distribution with parameters $r$ and
$k=(m-1)r$.

Namely, again let $V_1,\ldots,V_m$ be the total precipitation
volumes during $m$ wet periods and, moreover, $V_1\geqslant V_j$ for all
$j\geqslant 2$. Calculate the quantity
\begin{equation*}
SR_0=\frac{(m-1)V_1}{V_2+\ldots+V_m}
\end{equation*}
($SR_0$ means ``$S$ample $R_0$''). From what was said above it
follows that under the hypothesis $H_0$: ``the precipitation volume
$V_1$ under consideration {\em is not} anomalously large'' the r.v.
$SR$ has the Snedecor--Fisher distribution with parameters $r$ и
$k=(m-1)r$. Let $\varepsilon\in(0,1)$ be a small number,
$q_{r,k}(1-\varepsilon)$ be the $(1-\varepsilon)$-quantile of the
Snedecor--Fisher distribution with parameters $r$ и $k=(m-1)r$. If
$SR_0>q_{r,k}(1-\varepsilon)$, then the hypothesis $H_0$ must be
rejected, that is, the volume $V_1$ of precipitation during one wet
period must be regarded as anomalously large. Moreover, the
probability of erroneous rejection of $H_0$ is equal to
$\varepsilon$.

Let $l$ be a natural number, $1\leqslant l<m$. It is worth noting that,
unlike the test based on the statistic $R$, the test based on $R_0$
can be modified for testing the hypothesis $H_0'$: ``the
precipitation volumes $V_{i_1},V_{i_2},\ldots,V_{i_l}$ {\em do not}
make an anomalously large cumulative contribution to the total
precipitation volume $V_1+\ldots+V_m$''. For this purpose denote
\begin{equation*}
T_l=V_{i_1}+V_{i_2}+\ldots+V_{i_l},\ \ \ \ T=V_1+V_2+\ldots+V_m
\end{equation*}
and consider the quantity
\begin{equation*}
SR_0'=\frac{(m-l)T_l}{l(T-T_l)}.
\end{equation*}
In the same way as it was done above, it is easy to make sure that
\begin{equation*}
SR_0'\eqd\frac{(m-l)G_{lr,l}}{lG_{(m-l)r,1}}\eqd Q_{lr,(m-l)r}.
\end{equation*}

Let $\varepsilon\in(0,1)$ be a small number,
$q_{lr,(m-1)r}(1-\varepsilon)$ be the $(1-\varepsilon)$-quantile of
the Snedecor--Fisher distribution with parameters $lr$ и $k=(m-l)r$.
If $SR_0'>q_{lr,(m-l)r}(1-\varepsilon)$, then the hypothesis $H_0'$
must be rejected, that is, the cumulative contribution of the
precipitation volumes $V_{i_1},V_{i_2},\ldots,V_{i_l}$ into the
total precipitation volume $V_1+\ldots+V_m$ must be regarded as
anomalously large. Moreover, the probability of erroneous rejection
of $H_0'$ is equal to $\varepsilon$.

The examples of application of the test for a total precipitation
volume within a wet period to be anomalously large will be discussed
in Section~\ref{SecAnalysysVol}.

\subsection{Determination of abnormalities types based on the results of the statistical analysis}
\label{SecAbnorm}

In this section we present the results of the application of the
test $SR_0$ to the analysis of the time series of daily
precipitation observed in Potsdam and Elista from $1950$ to $2009$.

First of all it should be emphasized that the parameter $m$ of the
Snedecor--Fisher distribution of the test statistic $SR_0$ is
tightly connected with the time horizon, the abnormality of
precipitation within which is studied. Indeed, the average duration
of a wet/dry period (or, which is the same, the average distance
between the first days of successive wet periods) in Potsdam turns
out to be $5.804\approx 6$ days. So, one observation of a total
precipitation during a wet period, on the average, corresponds to
approximately 6 days. This means, that, for example, the value $m=5$
corresponds to approximately one month on the time axis, the value
$m=15$ corresponds to approximately 3 months (a season), the value
$m=60$ corresponds to approximately one year.

Second, it is important that the test for a total precipitation
volume during one wet period to be anomalously large can be applied
to the observed time series in a moving mode. For this purpose a
{\em window} (a set of successive observations) should be
determined. The number of observations in this set, say, $m$, is
called the {\em window width}. The observations within a window
constitute the sample to be analyzed. After the test has been
performed for a given position of the window, the window moves
rightward by one observation so that the leftmost observation at the
previous position of the window is excluded from the sample and the
observation next to the rightmost observation is added to the
sample. The test is performed once more and so on. It is clear that
each fixed observation falls in exactly $m$ successive windows (from $m$th to $N-m+1$, where $N$ denotes the number of wet periods). Two cases are possible: (i) the fixed observation is recognized as
anomalously large within {\em each} of $m$ windows containing this
observation and (ii) the fixed observation is recognized as
anomalously large within {\em at least one} of $m$ windows containing
this observation. In the case (i) the observation will be called
{\em absolutely anomalously large} with respect to a given time
horizon (approximately equal to $m\cdot 5.804\approx 6m$ days). In
the case (ii) the observation will be called {\em relatively
anomalously large} with respect to a given time horizon.
Of course, these definitions admit intermediate cases where the
observation is recognized as anomalously large for $q\cdot m$ windows
with $q\in[\frac{1}{m},1]$.

\subsection{The examples of statistical analysis of total precipitation volumes}
\label{SecAnalysysVol}

\begin{figure}[!h]
\begin{minipage}[h]{\textwidth}
\center{\includegraphics[width=\textwidth,
height=0.2\textheight]{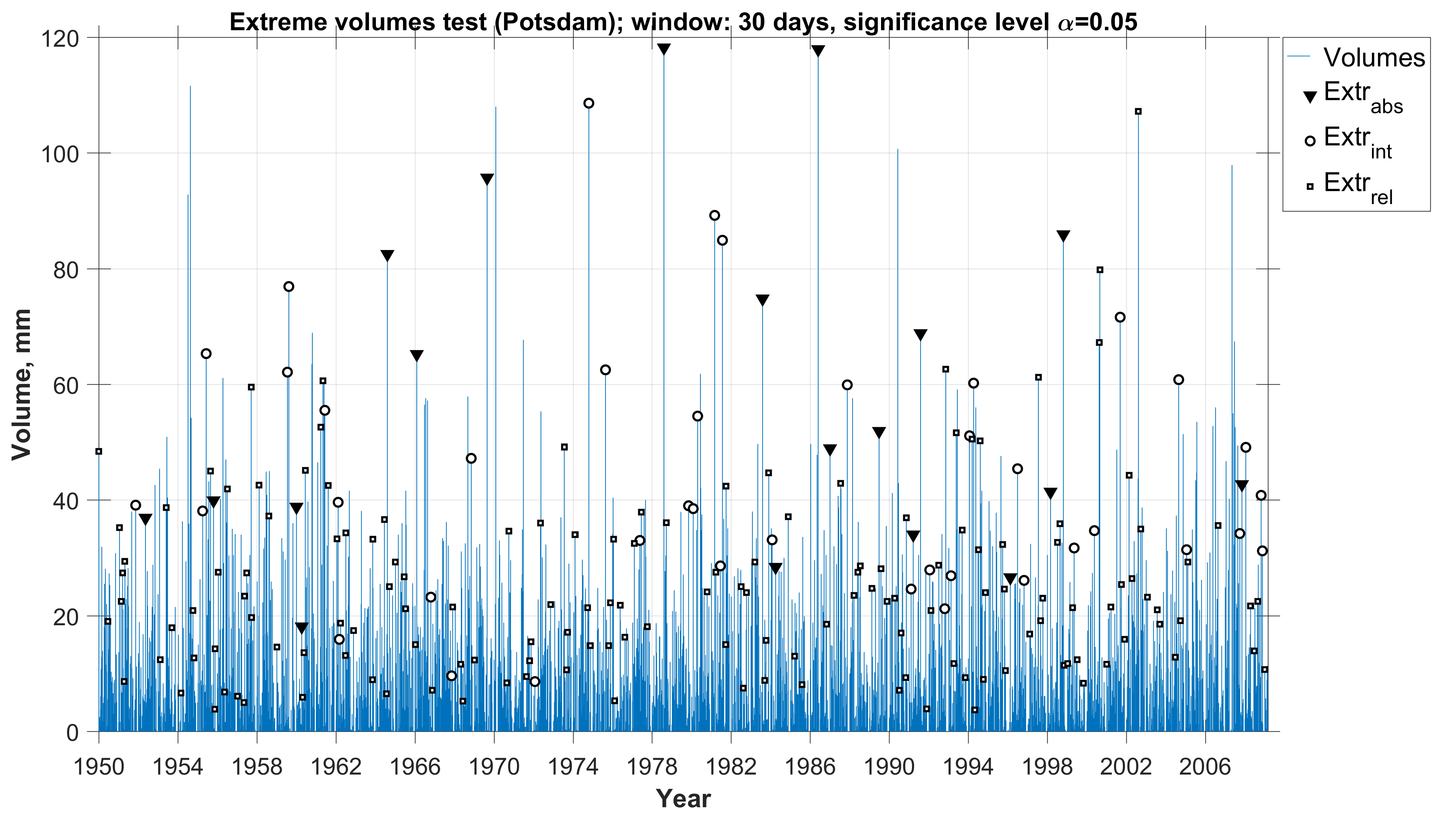} a)}
\end{minipage}
\hfill
\begin{minipage}[h]{\textwidth} \center{
\includegraphics[width=\textwidth,
height=0.2\textheight]{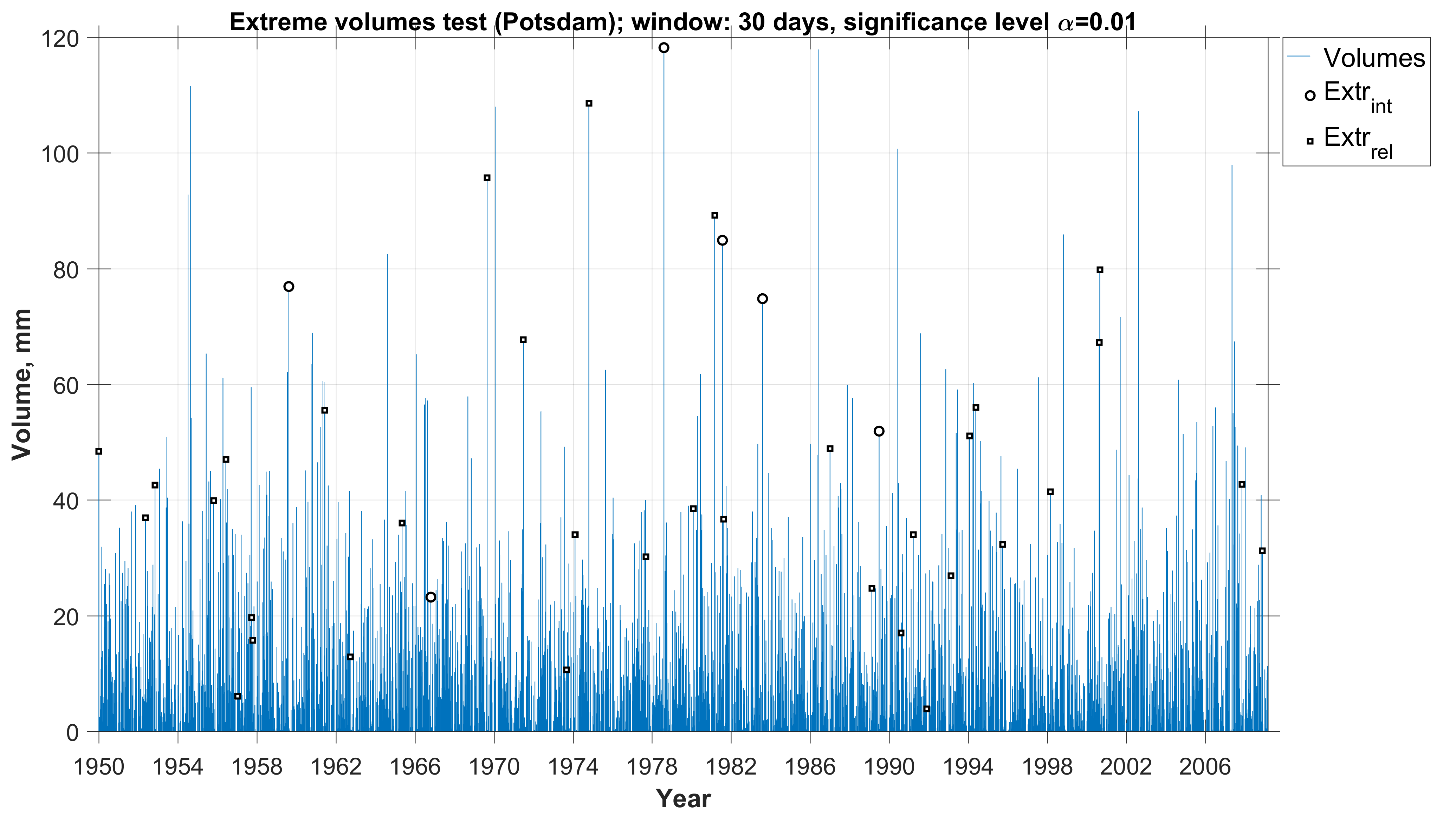} b)}
\end{minipage}
\caption{Absolutely (triangles), relatively
(squares) and intermediate (circles) abnormal precipitation volumes,
Potsdam, time horizon = 30 days, significance levels
$\alpha=0.05$ (a) and $\alpha=0.01$ (b).}
\label{FigPotsdamVolExtrTest30}
\end{figure}

\begin{figure}[!h]
\begin{minipage}[h]{\textwidth}
\center{\includegraphics[width=\textwidth,
height=0.2\textheight]{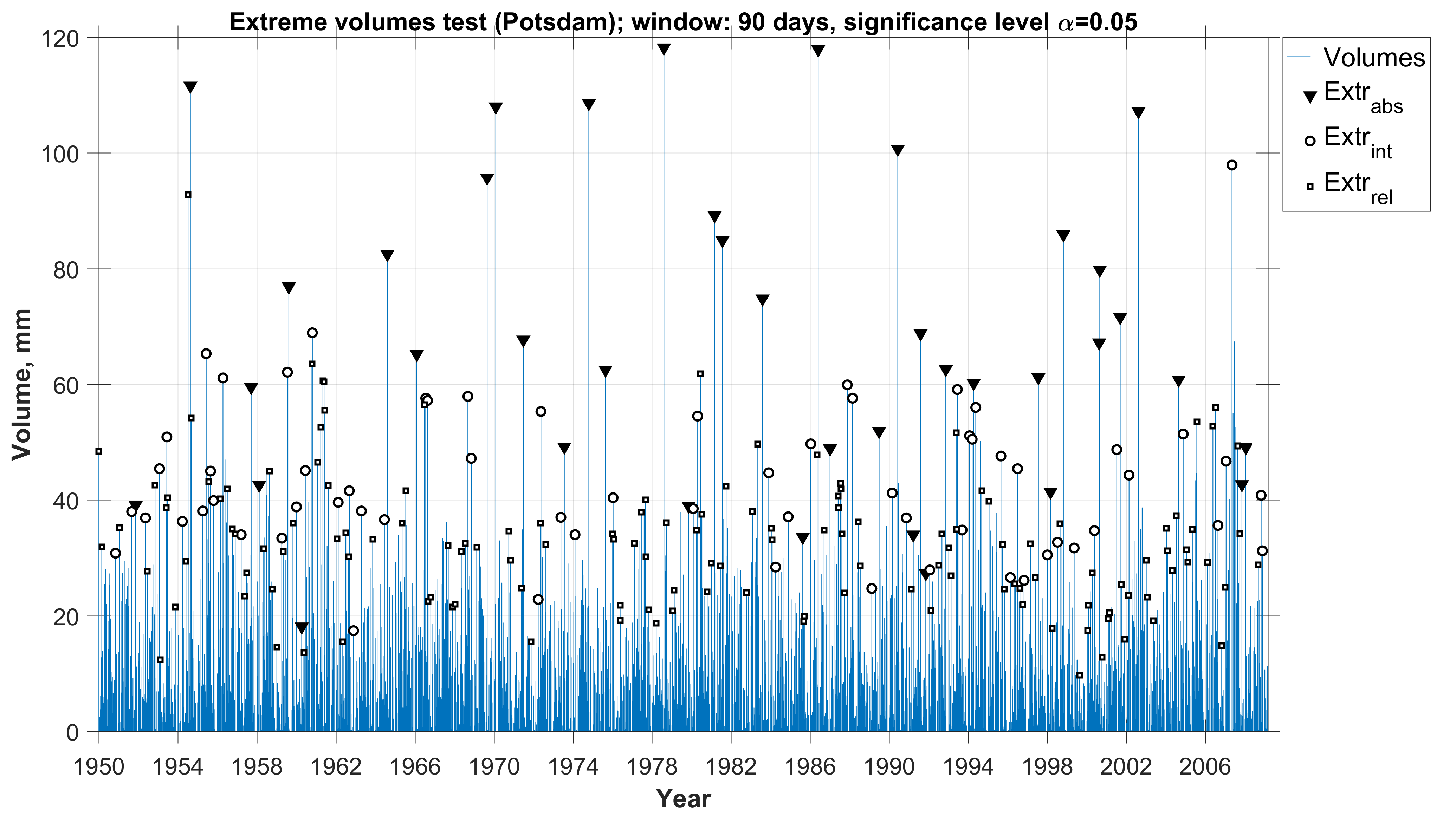} a)}
\end{minipage}
\vfill
\begin{minipage}[h]{\textwidth} \center{
\includegraphics[width=\textwidth,
height=0.2\textheight]{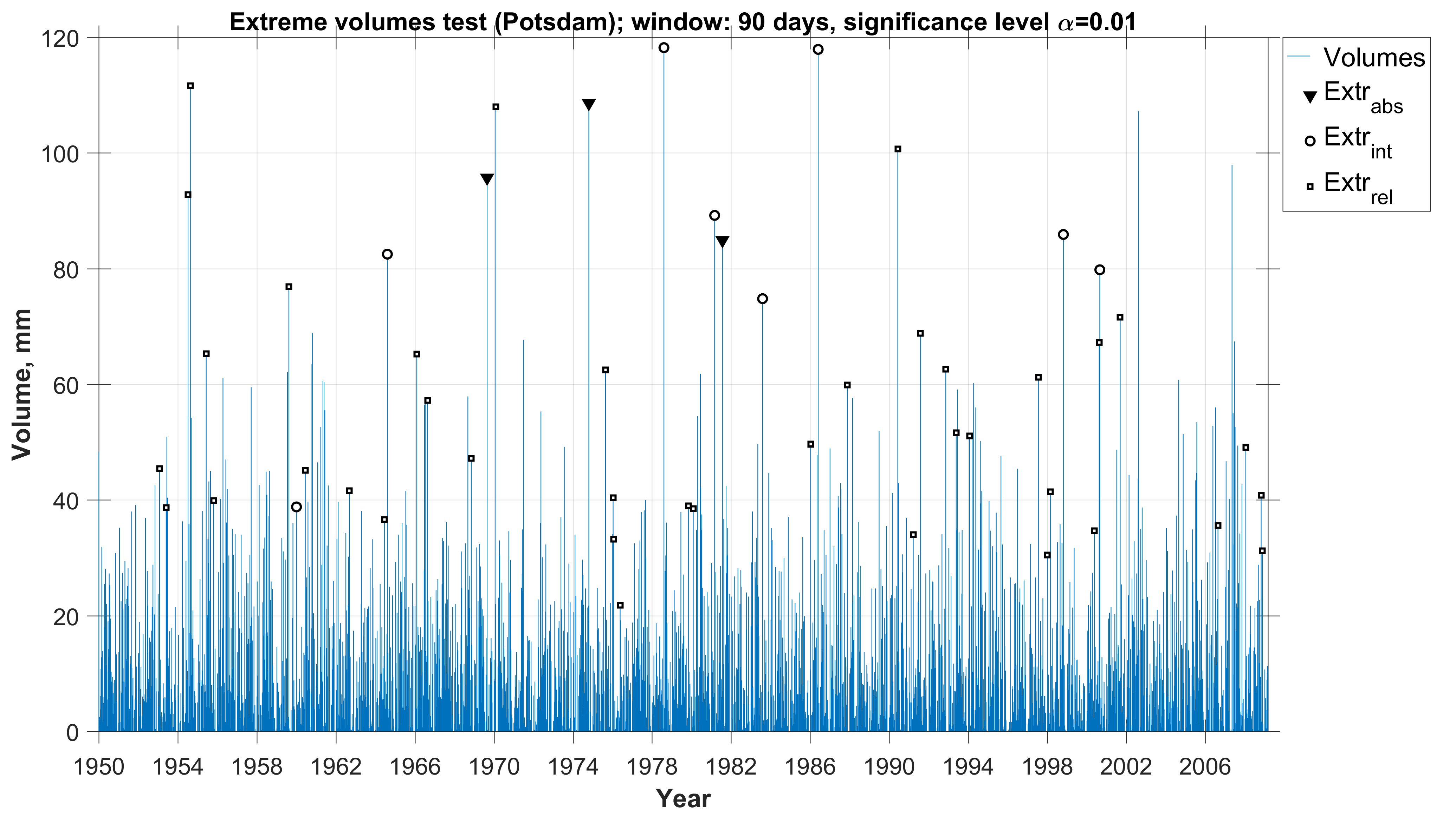} b)}
\end{minipage}
\caption{Absolutely (triangles), relatively
(squares) and intermediate (circles) abnormal precipitation volumes,
Potsdam, time horizon = 90 days, significance levels
$\alpha=0.05$ (a) and $\alpha=0.01$ (b).}
\label{FigPotsdamVolExtrTest90}
\end{figure}

\begin{figure}[!h]
\begin{minipage}[h]{\textwidth}
\center{\includegraphics[width=\textwidth,
height=0.2\textheight]{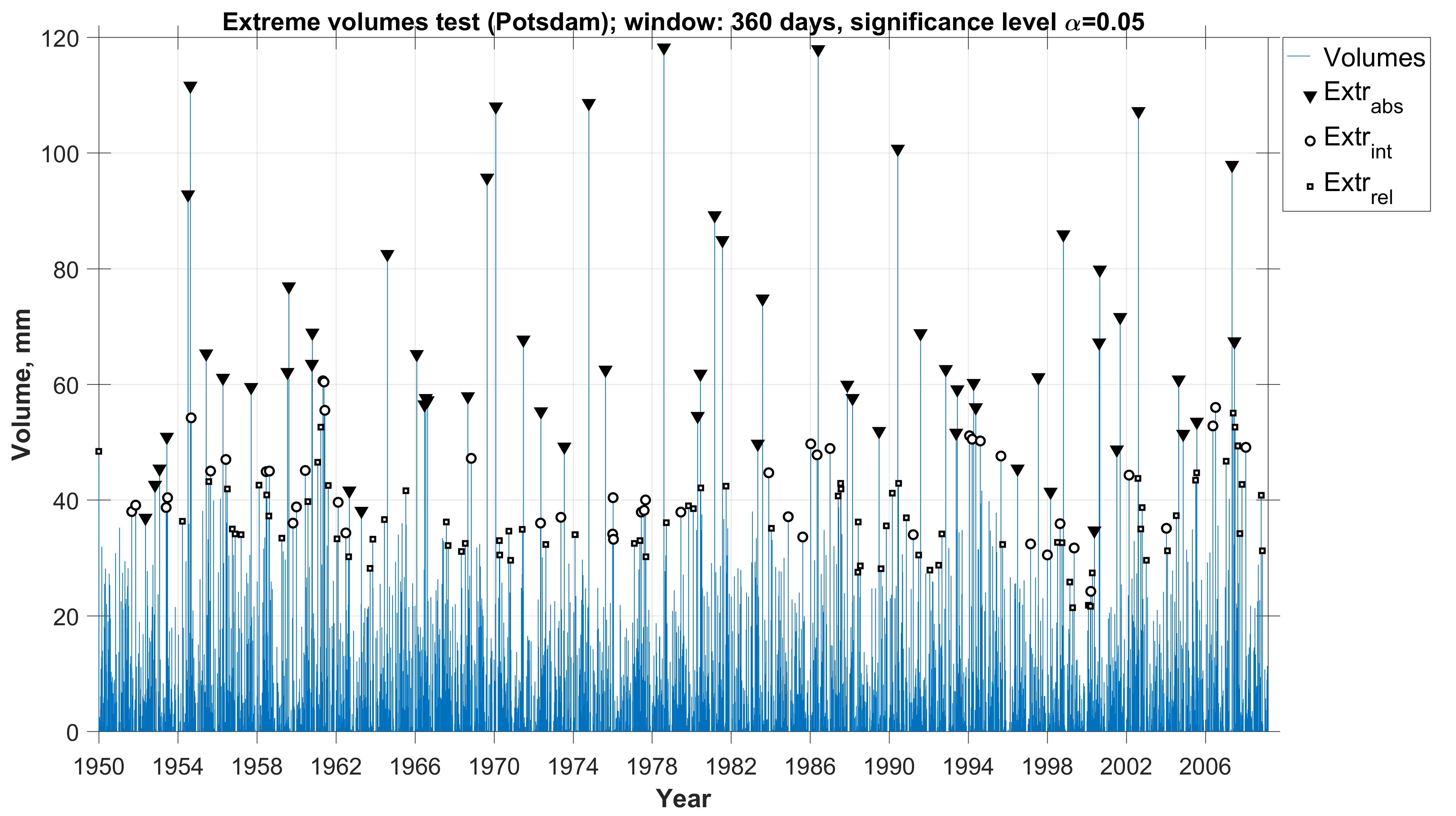} a)}
\end{minipage}
\vfill
\begin{minipage}[h]{\textwidth} \center{
\includegraphics[width=\textwidth,
height=0.2\textheight]{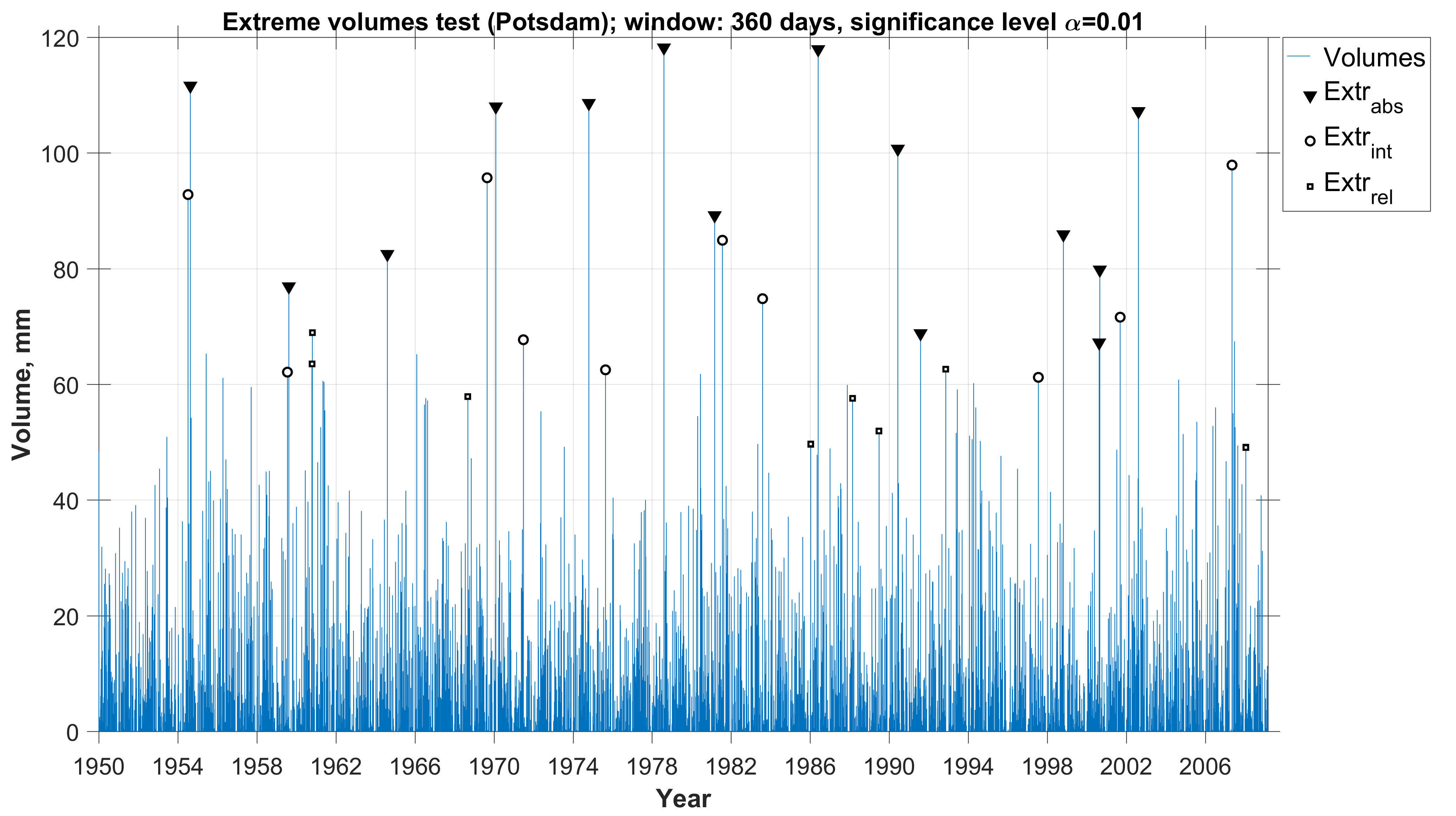} b)}
\end{minipage}
 \caption{Absolutely (triangles), relatively
(squares) and intermediate (circles) abnormal precipitation volumes,
Potsdam, time horizon = 360 days, significance levels
$\alpha=0.05$ (a) and $\alpha=0.01$ (b).}
\label{FigPotsdamVolExtrTest360}
\end{figure}

\begin{figure}[!h]
\begin{minipage}[h]{\textwidth}
\center{\includegraphics[width=\textwidth,
height=0.2\textheight]{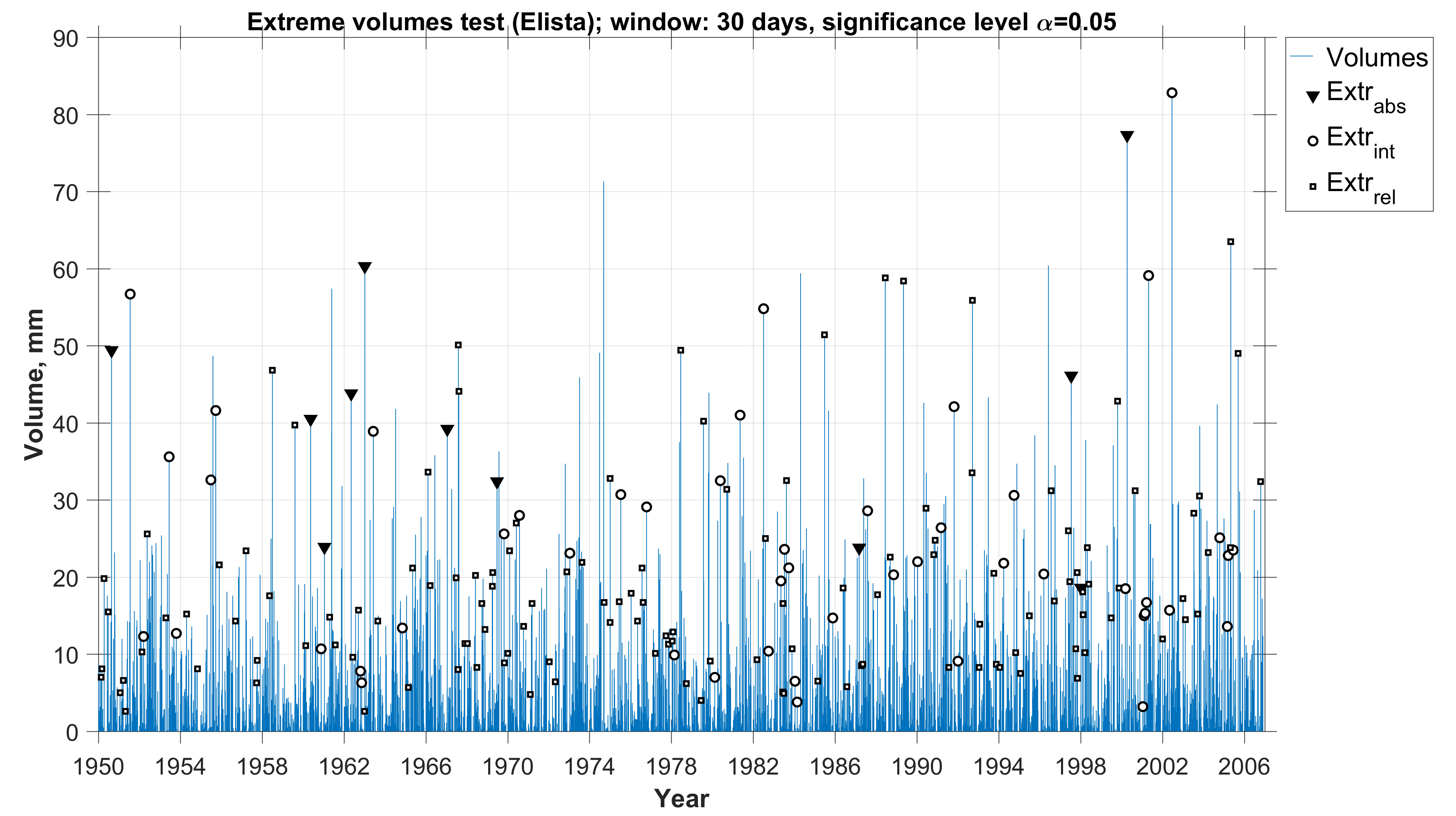} a)}
\end{minipage}
\vfill
\begin{minipage}[h]{\textwidth} \center{
\includegraphics[width=\textwidth,
height=0.2\textheight]{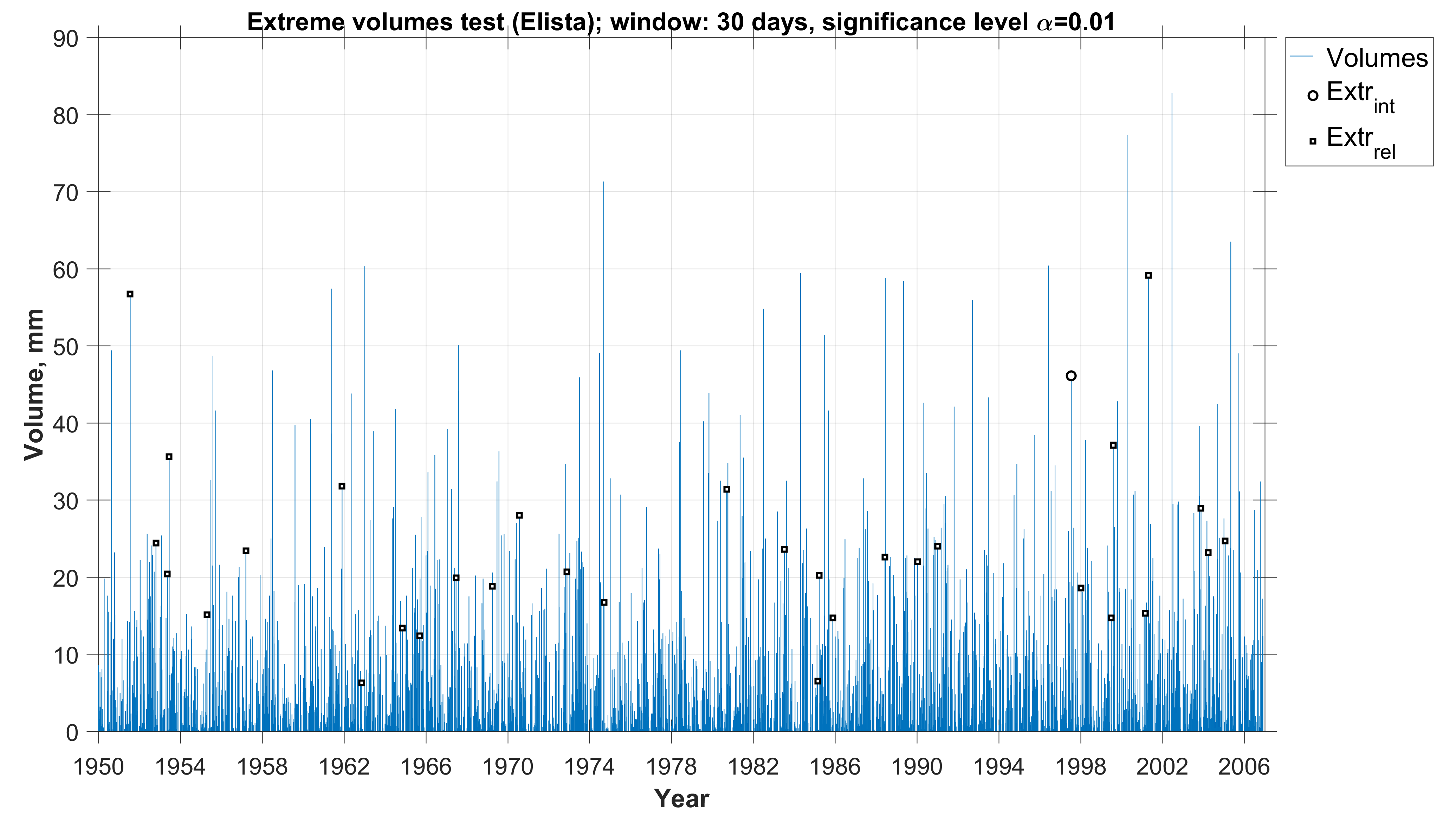} b)}
\end{minipage}
 \caption{Absolutely (triangles), relatively
(squares) and intermediate (circles) abnormal precipitation volumes,
Elista, time horizon = 30 days, significance levels
$\alpha=0.05$ (a) and $\alpha=0.01$ (b).}
\label{FigElistaVolExtrTest30}
\end{figure}

\begin{figure}[!h]
\begin{minipage}[h]{\textwidth}
\center{\includegraphics[width=\textwidth,
height=0.2\textheight]{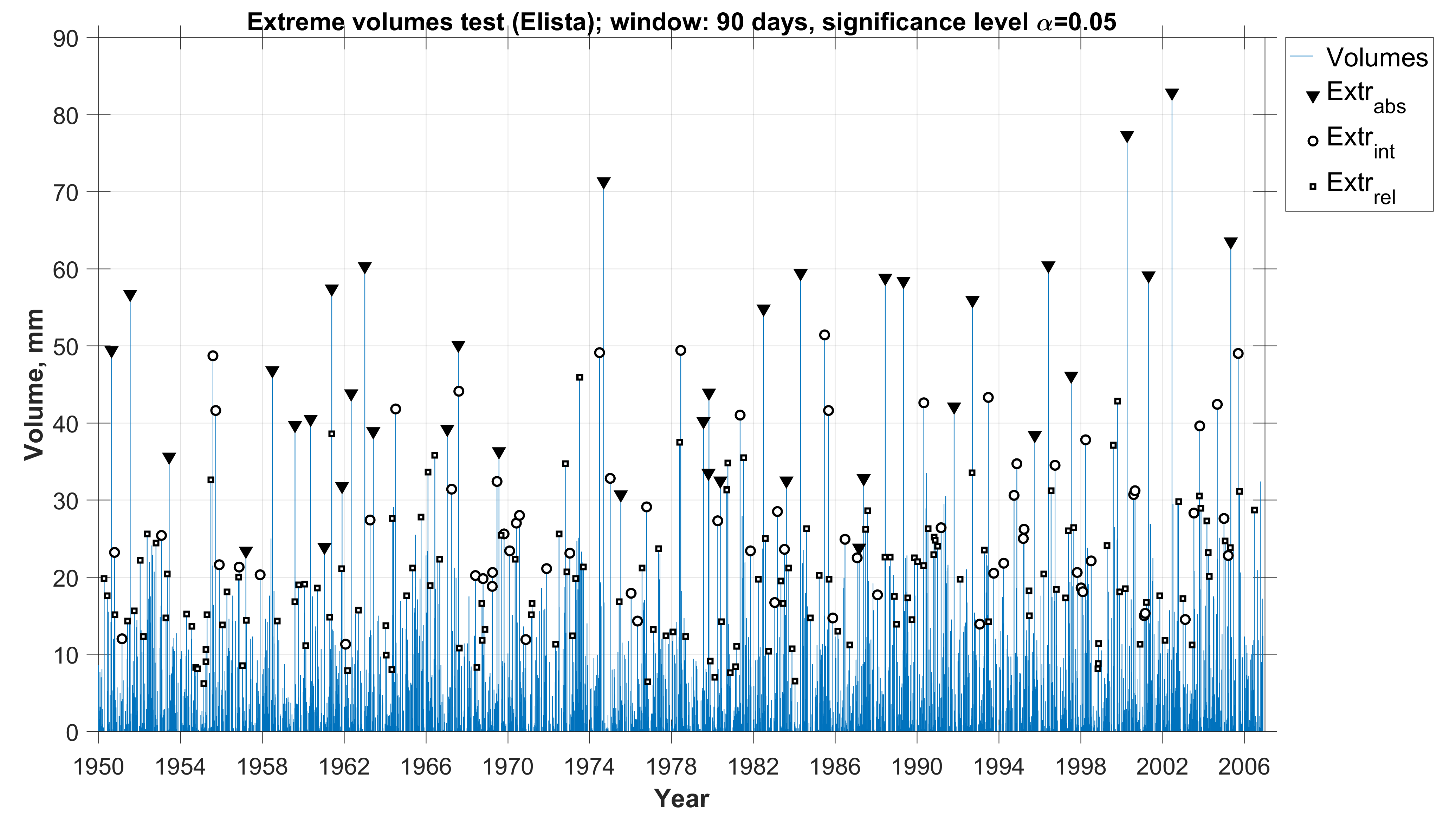} a)}
\end{minipage}
\vfill
\begin{minipage}[h]{\textwidth} \center{
\includegraphics[width=\textwidth,
height=0.2\textheight]{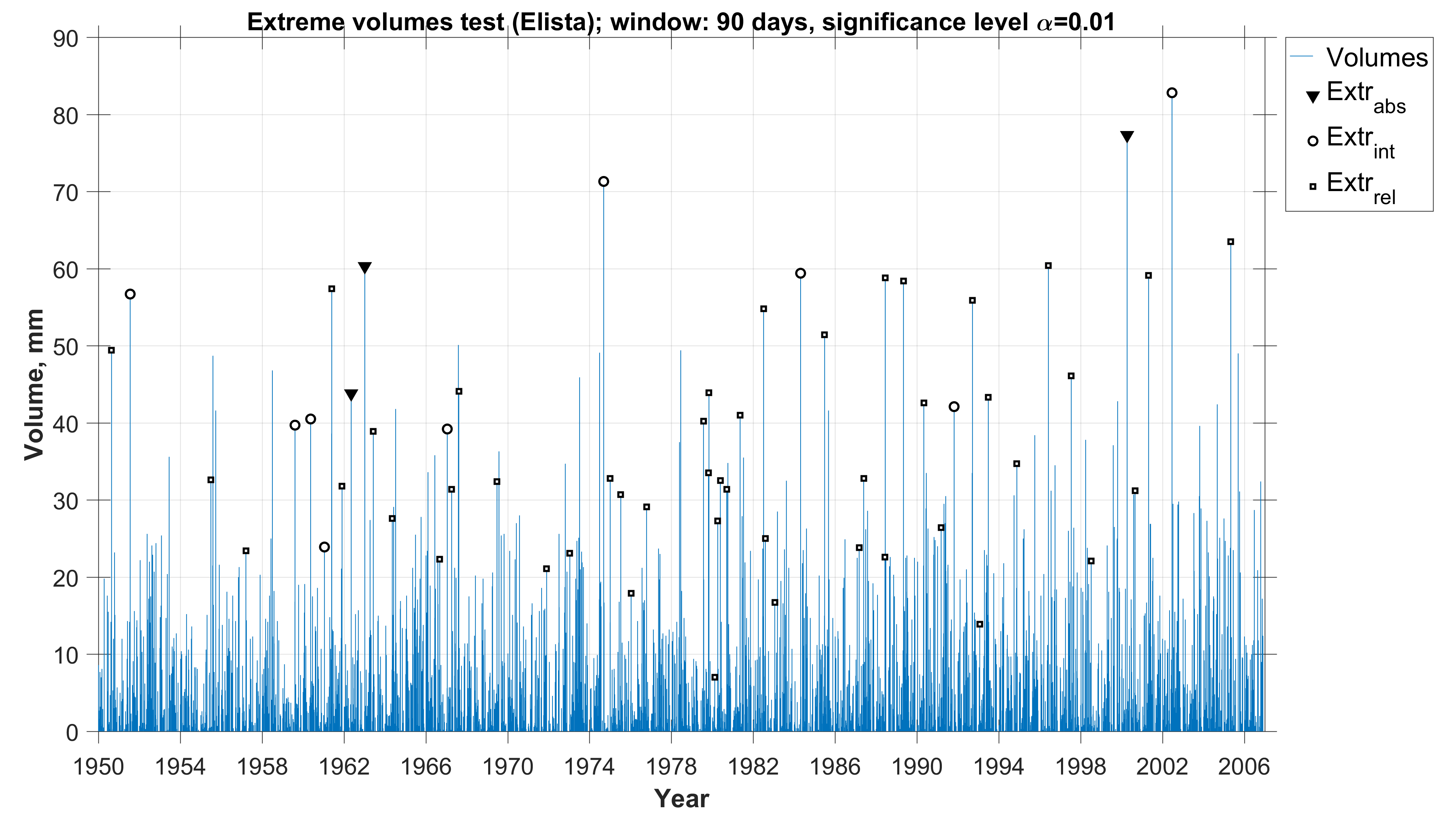} b)}
\end{minipage}
 \caption{Absolutely (triangles), relatively
(squares) and intermediate (circles) abnormal precipitation volumes,
Elista, time horizon = 90 days, significance levels
$\alpha=0.05$ (a) and $\alpha=0.01$ (b).}
\label{FigElistaVolExtrTest90}
\end{figure}

\begin{figure}[!h]
\begin{minipage}[h]{\textwidth}
\center{\includegraphics[width=\textwidth,
height=0.2\textheight]{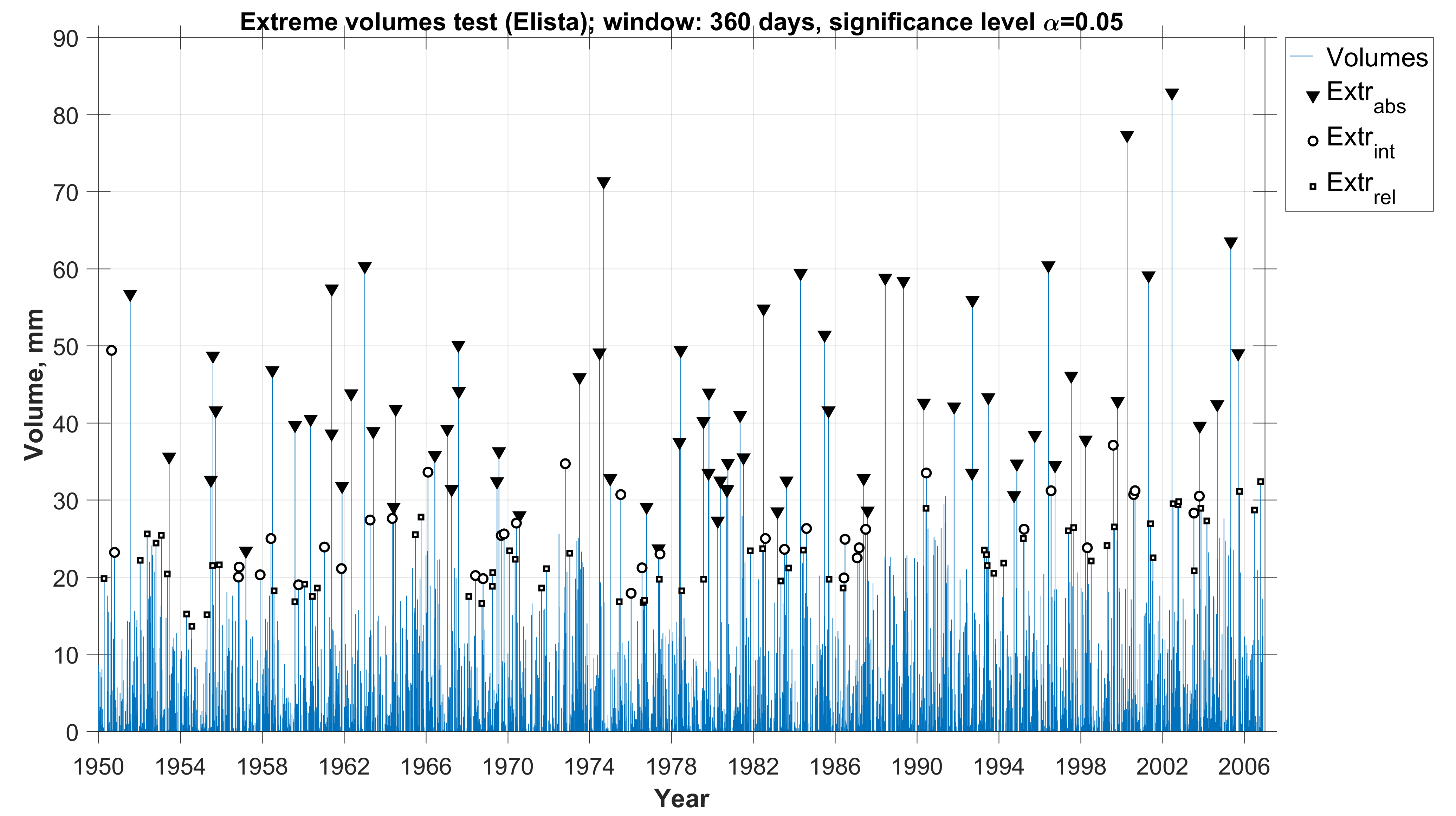} a)}
\end{minipage}
\vfill
\begin{minipage}[h]{\textwidth} \center{
\includegraphics[width=\textwidth,
height=0.2\textheight]{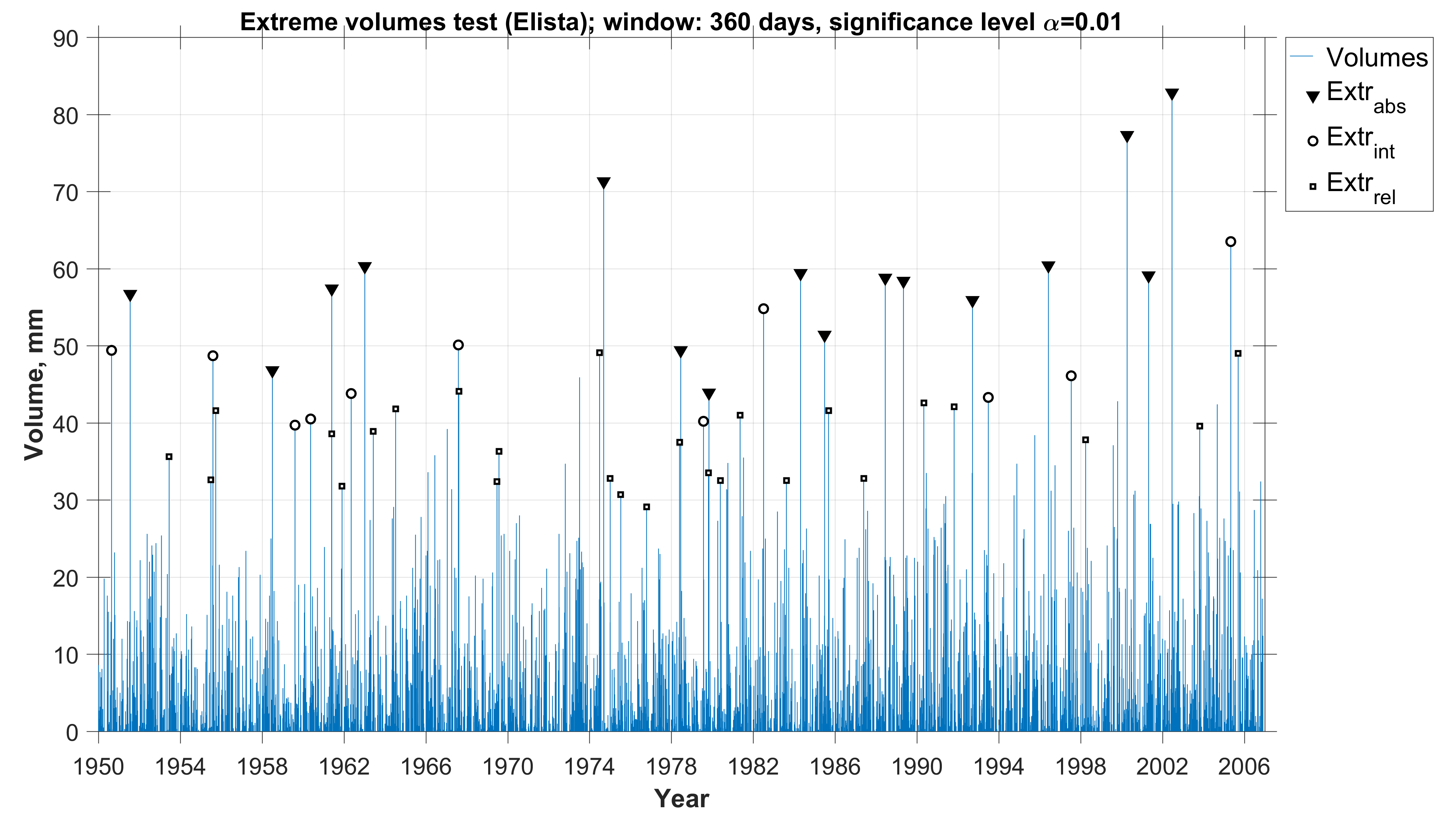} b)}
\end{minipage}
 \caption{Absolutely (triangles), relatively
(squares) and intermediate (circles) abnormal precipitation volumes,
Elista, time horizon = 360 days, significance levels
$\alpha=0.05$ (a) and $\alpha=0.01$ (b).}
\label{FigElistaVolExtrTest360}
\end{figure}

The results of the application of the test for a total precipitation
volume during one wet period to be anomalously large based on $SR_0$
in the moving mode are shown on Figs.~\ref{FigPotsdamVolExtrTest30}--\ref{FigPotsdamVolExtrTest360} (Potsdam) and \ref{FigElistaVolExtrTest30}--\ref{FigElistaVolExtrTest360} (Elista) for different time horizons ($30$, $90$ and $360$ days). The notation $Extr_{int}$ corresponds to the {\em intermediate} extremes (the fixed observation is recognized as anomalously large within at least $\lceil m/2\rceil$ windows containing this observation, here the symbol $\lceil\cdot \rceil$ denotes the next larger integer).

It is seen that at relatively small time horizons the test yields non-trivial and
unobvious conclusions. However, as the time horizon increases, the
results of the test become more expected. At small time horizons
there are some big precipitation volumes that are not recognized as
abnormal. At large time horizons there are almost no `regular' big
precipitation volumes at significance level $\alpha=0.05$
whereas at the smaller significance level $\alpha=0.01$ there
are some `regular' big precipitation volumes which are thus not
recognized as abnormal.

\section{Conclusions and discussion}
\label{SecConclusion}
This paper states that the negative binomial distribution may be fruitful for description of the really observed wet periods and can be applied to test of hypothesis that the specific precipitation volume considered over given wet period is anomalous. It is an important issue since now there is no single-valued criterion which precipitation volume is anomalous and which is not. Obviously that the same volume may be normal in one region where precipitations are quite frequent, for instance in tropical zone and absolutely anomalous in another one, for instance in desert. The proposed test considers the relative part of precipitation and is free from the aforementioned disadvantage. On the other hand, it gives the numerical method how to test this hypothesis that can be easily implemented.
The considered scheme may be expanded for other geophysical variables such as wind speed, heat fluxes etc, both separately or jointly. This may be very important for global climate prediction models, for forecasting and evaluation of dangerous phenomena and processes.

\section*{Acknowledgements}
The research was partially supported by the {\color{blue}Russian Foundation for Basic Research} (project~{\color{blue}17-07-00851)} and the {\color{blue}RF Presidential scholarship program} (No.~{\color{blue}538.2018.5}).

\end{document}